# Title: Growth Model Interpretation of Planet Size Distribution

**Authors:** Li Zeng*[1,2], Stein B. Jacobsen[1], Dimitar D. Sasselov[2], Michail I. Petaev[1,2], Andrew Vanderburg[3], Mercedes Lopez-Morales[2], Juan Perez-Mercader[1], Thomas R. Mattsson[4], Gongjie Li[5], Matthew Z. Heising[2], Aldo S. Bonomo[6], Mario Damasso[6], Travis A. Berger[7], Hao Cao[1], Amit Levi[2], Robin D. Wordsworth[1].

**Affiliations:**
[1]Department of Earth and Planetary Sciences, Harvard University, 20 Oxford Street, Cambridge, MA 02138
[2]Center for Astrophysics | Harvard & Smithsonian / Department of Astronomy, Harvard University, 60 Garden Street, Cambridge, MA 02138
[3]Department of Astronomy, The University of Texas at Austin, Austin, TX 78712
[4]High Energy Density Physics Theory Department, Sandia National Laboratories, PO Box 5800 MS 1189, Albuquerque, NM 87185
[5]School of Physics, Georgia Institute of Technology, Atlanta, GA 30313
[6]INAF-Osservatorio Astrofisico di Torino, via Osservatorio 20, 10025 Pino Torinese, Italy
[7]Institute for Astronomy, University of Hawaii, 2680 Woodlawn Drive, Honolulu, HI 96822

*Correspondence to: astrozeng@gmail.com

**Abstract: The radii and orbital periods of 4000+ confirmed/candidate exoplanets have been precisely measured by the *Kepler* mission. The radii show a bimodal distribution, with two peaks corresponding to smaller planets (likely rocky) and larger intermediate-size planets, respectively. While only the masses of the planets orbiting the brightest stars can be determined by ground-based spectroscopic observations, these observations allow calculation of their average densities placing constraints on the bulk compositions and internal structures. Yet an important question about the composition of planets ranging from 2 to 4 Earth radii ($R_\oplus$) still remains. They may either have a rocky core enveloped in a $H_2$-He gaseous envelope (gas dwarfs) or contain a significant amount of multi-component, $H_2O$-dominated ices/fluids (water worlds). Planets in the mass range of 10-15 $M_\oplus$, if half-ice and half-rock by mass, have radii of 2.5 $R_\oplus$, which exactly match the second peak of the exoplanet radius bimodal distribution. Any planet in the 2-4 $R_\oplus$ range requires a gas envelope of at most a few mass%, regardless of the core composition. To resolve the ambiguity of internal compositions, we use a growth model and conduct Monte Carlo simulations to demonstrate that many intermediate-size planets are "water worlds".**

**Significance Statement**: The discovery of numerous exoplanet systems containing diverse populations of planets orbiting very close to their host stars challenges the planet formation theories based on the Solar system. Here we focus on the planets with radii of 2-4 $R_\oplus$, whose compositions are debated. They are thought to be either gas dwarfs consisting of rocky cores embedded in $H_2$-rich gas envelopes or water worlds containing significant amounts of $H_2O$-dominated fluid/ice in addition to rock and gas. We argue that these planets are water worlds.

**\body**

Thousands of exoplanets discovered during the last two decades cover a wide range of masses and sizes. In the 1-20 $M_\oplus$ and 1-4 $R_\oplus$ ranges several types of planets have been identified. The planets with high densities are considered rocky, while the ones with low densities have been modeled either as gas dwarfs or water worlds. The gas dwarfs are thought to have a rocky core enveloped in a $H_2$-He gaseous envelope while the water worlds contain a significant amount of multi-component, $H_2O$-dominated ices/fluids in addition to rock and gas.

Here we focus on such small planets, namely the super-Earths (1-2 $R_\oplus$) and the sub-Neptunes (2-4 $R_\oplus$). Mass-radius curves (**Figs. 1 & 2**) show that they contain a few percent of gas by mass at most; *i.e.*, their masses are dominated by the cores that must have formed by the accretion of solids in the disk.

The measurements of planetary radii and orbital periods of more than 4000 confirmed or candidate exoplanets by the NASA *Kepler* Mission (*1–6*) revealed a bimodal distribution of planet sizes in the 1-4 $R_\oplus$ range. This discovery was interpreted as the evidence for the presence of two populations of planets – smaller rocky worlds and the intermediate-size planets, which, because of their proximity to the host star, were interpreted as gas dwarfs rather than water worlds.

Further refinement of the host stellar parameters by the Gaia astrometry mission (*7–11*) yields a better resolved bimodal distribution of planetary radii with two peaks and a local minimum or gap at 1.8-2 $R_\oplus$ (**Figs. 1 & 2**). The gap separates two sub-populations of planets on the mass-radius diagram (**Figs. 1 & 2**): super-Earths (1-2 $R_\oplus$) and sub-Neptunes (2-4 $R_\oplus$). Whether this gap is a direct result of planet formation or a secondary feature formed by photo-evaporation of gas envelopes is debated. The distances to the host stars in both sub-populations appear to be distributed log-uniform, i.e., flat in semi-major axis **a** or orbital period **P**, beyond ~10 days (Fig. 7 of (*12*)). Such orbit distributions of both populations challenge the photo-evaporation scenario as the cause of the gap because it strongly depends on the orbital distance. If the gap is caused by photo-evaporation, then an anti-correlation in their orbital distributions is expected (*13*). We interpret the flatness in distributions as additional evidence for both populations arising from their intrinsic properties (*14*).

The distinction between gas dwarfs and water worlds cannot be made based on the mass-radius relationship alone or the presence or lack of a $H_2$-He gas layer (*15*). Therefore, we invoke a mass-radius distribution among small planets and their growth models to argue that at least some intermediate-size planets are water worlds.

**Protoplanetary disks** of solar-like composition contain three principal planet-building components – nebular gas, $H_2O$-rich ices, and rocky (silicates+Fe,Ni metal) materials - whose compositions are a function of element volatility (characterized by the equilibrium condensation temperatures, **Fig. 3**), and, therefore, change radially and temporally with changing temperature. In the case of cooling, gradual decrease in temperature results in progressive condensation of Fe-Mg-Ca-Al silicates and Fe,Ni-metal making up 0.5% of the total disk mass at ~1000 K, until C,N,O,H-bearing ices (1.0% total mass) start to sequentially condense below ~200 K. The condensation of ices is a sharp feature in the protoplanetary disk. Once the temperature drops a bit (a few Kelvins) below its condensation temperature, a very large amount of ice would form. In the case of heating, *e.g.* due to inward dust drift, the sequence of phase changes reverses. The disk never gets cold enough for condensation of He and $H_2$ so ~98 wt.% of the disk always remains gaseous.

Because the condensation temperatures of silicate and ice phases are quite different, it is their condensation fronts in the mid-plane of protoplanetary disks that set the rough boundaries of planetesimals with different compositions (*16*). The most important among them – the snowline – marks the stability field of the $H_2O$ ice. The presence of ices would significantly enhance the local mass surface density of solids in the disk. $H_2O$ snowlines are prominent features in protoplanetary disks, which have been predicted theoretically (*17–19*), then inferred around TW Hya (*20*) and HL Tau (*21*), and now observed by ALMA around young star V883 Ori undergoing FU Ori outburst which pushes the water snowline to 40 a.u. to make it observable (*22*, *23*).

If a planet forms in the presence of ices, the phase diagram predicts a similar amount of multi-component, $H_2O$-dominated ices to be added to the rocky material. The rock/ice ratio does not strongly depend on the host star metallicity ([Fe/H] or [M/H]) because the metallicity mostly reflects the ratio of the total condensable solid materials (metal+rock+ices) to the $H_2$+He gas.

**Planet formation** in the Solar System is thought to have started in an accretion disk that fed the initial mixture of interstellar $H_2$+He gas, C,N,O-rich ices and Mg,Si,Fe-rich silicates to the growing Sun. At some point, the disk became thermally zoned, with the inner regions being hot enough for complete evaporation of all ices and some silicates and the colder outer zone where only a portion of ices could evaporate. Within each zone the dust grains first coagulated into kilometer-sized planetesimals, then, within ~$10^5$ years, Moon- to Mars-sized planetary embryos accreted. The initial differences in radial proportions of silicates and ices in the protoplanetary disk, along with other factors, resulted in the formation of three types of planets in our solar system – the small terrestrial rocky planets, the large gas-giants Jupiter and Saturn, and the intermediate ice-giants Uranus and Neptune. Whether the formation of our Solar System planets is typical or not is still an open question.

**Pebble accretion** has been proposed to explain the fast growth of planets to a few Earth masses ((*24*), Fig.7). Growth by pebble accretion is only effective for icy pebbles where $H_2O$ occurs as layers of ice coating silicate dust (*25*). When drifting across the snowline, ices may (partially) sublimate (**SI Appendix, Fig. S1**). The inward drift of icy pebbles may effectively push the snowline closer to the star (*26*), making the snowline a dynamic feature with location changing as the disk evolves (*27–29*). If a planet growth involves pebble accretion then, regardless of the detailed growth mechanisms, both ice and rock should participate in the growth process, with the ice amount being comparable to the silicates. In the case of gas dwarfs this means that at least some of their cores should contain ices in addition to rock.

Icy planets can migrate inward through planet-disk interactions (*30*) or planet-planet scattering (*31*) and continue growing, as illustrated by the growth tracks in the planet mass-semi major axis diagrams of (*32*). We propose such a mechanism to form the two sub-populations of exoplanets with radii between 1-4 $R_\oplus$. This point is reinforced by the bimodality of densities, and thus, compositions, of the satellites of Jupiter, Saturn, Uranus, Neptune (**SI Appendix, Fig. S2**).

To interpret the observed mass-radius distributions we modeled the growth curves of planets by adding either ice or gas to a rocky core (see **Materials and Methods** and the **SI appendix**).

The **mass distributions** of small planets are compared with each other (**Fig. 2,** histograms on the X-axis) and the calculated growth tracks (**Fig.1**). Planets of 1.4-1.9 $R_\oplus$ and 2-3 $R_\oplus$ are compared, where each bin contains about 20 planets (**SI Appendix, Table S1**). The mass distributions of sub-Neptunes range from 3-20 $M_\oplus$, while super-Earths truncate at 10 $M_\oplus$. The first population has masses of 5.3±1.5 $M_\oplus$. The second population has masses of 8.3±3 $M_\oplus$. The offset of 3 $M_\oplus$ suggests that the cores of the second population are significantly more massive, which coincides with adding ice to rock (**Fig. 1**, cyan arrows – growth curves of ices).

**Bimodal mass distribution** of the RV (radial-velocity) sub-Neptune population (2-4 $R_\oplus$) is identified, and the higher peak of which, with the mass of ~16 $M_\oplus$, may result from the merging of primary icy cores. This can explain why Uranus (*33*), Neptune, and similar planets did not accrete significant amount of $H_2$+He gas, despite their final mass appears to be greater than the critical core mass (*34*) for a run-away gas accretion. This means that the Uranus, Neptune and many other icy cores like them are mergers of many smaller, less-massive, primordial icy cores. Each primordial core is less massive than the critical core mass. The collision and merging of two primordial cores can partially remove their $H_2$/He envelopes (*35*), if such existed, while retaining most of their core masses. The colliding icy cores are more likely to stick and merge, yielding a core of doubled mass. In contrast, a collision of two rocky cores, each bigger than ~10 $M_\oplus$, tends to catastrophically disrupt them rather than to merge (*36*).

**TTV (transit-timing-variation) planets** (triangles in **Fig. 2**) of 2-4 $R_\oplus$, on the other hand, show systematically smaller masses. They are preferably found around metal-poor host stars ([Fe/H]<-0.3) (*37*), as opposed to RV sub-Neptunes which are found across a wide range of metallicity (*12*), suggesting somewhat different formation environments.

**Atmospheric Escape** has a correlation between the escape velocities and atmospheric compositions of objects in our solar system (**Fig. 4**). Applying the same physics to exoplanets, one can calculate, for a given surface temperature, what gaseous species a planetary atmosphere can hold. Super-Earths and sub-Neptunes have escape velocities on the order of ~20 km/s, marked by the contour of $M_p/R_p$=3 in the Earth units (**Figs. 1 & 2**). 20 km/s roughly corresponds to the thermal escape threshold of atomic hydrogen at 150 K, or molecular hydrogen ($H_2$) at 300 K, or helium at 600 K. Therefore, bodies with $M_p/R_p$<3 are susceptible to the escape of $H_2$ and He. The $H_2$ in their atmospheres/envelopes cannot last over billion-year timescale unless being continuously replenished by an underlying reservoir. It is possible that methane $CH_4$, ammonia $NH_3$ (*38–40*), and even $H_2$ (*41*) outgas gradually from an initial $H_2O$ reservoir to replenish a primary envelope, or to form a secondary one. During this process, He and heavier species are preferentially retained (*42*). There is a huge density contrast among planets residing around the He escape threshold (gray dashed line labeled 'Helium' in **Fig. 4)**. Some of such planets, e.g. WASP-107 b (*43*, *44*), WASP-69 b (*45*), HD 189733 b (*46*), and HAT-P-11 b (*47*, *48*), are observed to be surrounded by He-rich extended atmospheres suggestive of He escape from them. Because of similar temperatures and escape velocities (depth of surface gravity potential well), all planets residing there would be losing He (and $H_2$) at this moment, if they have any. We infer that the He escape threshold is the boundary separating the populations of puffy hot-Saturns and small (<4 $R_\oplus$) exoplanets. The latter likely possess higher mean molecular weight atmospheres with $H_2O$ in their deeper interior (above a few GPa) partitioned among condensed phases including fluids, insulating solids and superionic ices (*49*).

**Water worlds** have other observational evidence including: **i.** Spectroscopy of metal-polluted white dwarfs showing that some planetary debris accreted by white dwarfs are ice-rich (*50*, *51*); **ii.** Transmission spectroscopy of an inflated Saturn-mass exoplanet WASP-39b that shows strong water-absorption features corresponding to the estimated $H_2O/H$ ratio of 151×solar in the planet's atmosphere (*52*).

**Monte Carlo simulation** was used to show that the bimodal radius distribution (yellow histograms on the Y-axis in **Figs. 1 & 2**) could arise from the dichotomy of rocky and icy cores. It is performed with the following assumptions. The proto-planetary disk is assumed to have solar-like major element ratios (Fe : Mg : Si : O : C : N), inferred from the tight distribution of C/O and Mg/Si ratios around solar values in the main-sequence stars in the solar neighborhood (*53*). Besides the $H_2$-He gas, the disk contains roughly one part of Mg-silicate-rock+(Fe,Ni)-metal, one part of $H_2O$-ice, and one part of other ices (methane clathrates and ammonia hydrates, which are still dominated by the $H_2O$ in their compositions) by mass. Due to the disk temperature gradient and condensation temperature gap between the rock+metal and ice, the disk is thermally zoned into an inner zone with solid rock+metal dominating, and an outer zone with both rock+metal and ices (mostly $H_2O$) suspended in $H_2$+He gas.

The simulation reproduces the bimodal radius distribution (**Fig. 2**, blue vertical histogram). It also reproduces the mass-radius distribution of RV planets with their gap, groupings (**Fig. 2**, gray contours in the background), and mass offset. Planets >3 $R_\oplus$ generally require the presence of a gaseous envelope.

**Conclusions**

We divide exoplanets around sun-like stars into four main categories according to the cumulative planet radius distribution (*14*) and mass-radius diagram (**Figs. 1 & 2**):

**i)** **Rocky worlds (<2 $R_\oplus$)** consist primarily of Mg-silicate-rock and (Fe,Ni)-metal; they broadly follow the extrapolation of mass-radius relation of Earth and Venus.

**ii)** **Water worlds (2-4 $R_\oplus$)** contain significant amounts (>1/4, and possibly more than 1/2, by mass) of $H_2O$-dominated-ices in addition to rock. TTV planets in this radius range are consistent with a less massive core possessing a gaseous envelope; they tend to be found around metal-poor stars (*37*).

**iii)** **Transitional planets (4-10 $R_\oplus$)** are likely to be ice-rich with substantial gaseous envelopes ($\gtrsim$5-10% by mass). They are typically a few tens of $M_\oplus$ forming a bridge between small exoplanets and gas giants on the mass-radius diagram. The micro-lensing surveys (*54*, *55*) find some of them at a few a.u. distance.

**iv)** **Gas giants (>10 $R_\oplus$)** are dominated by $H_2$-He in the bulk composition and have masses and radii comparable to Jupiter.

Our own solar system planets fit into this classification. However, two puzzles remain unsolved: the compactness of many *Kepler* planetary systems compared to our own solar system and the lack of planets intermediate in size between Earth and Neptune in our own solar system. These two puzzles may be interrelated. Solving them is the key to understanding the unique initial conditions that form our own solar system. The abundance of these intermediate-size planets (water worlds) in our galaxy challenges us – to understand their formation, migration, interior structure, atmosphere, and habitability.

**Materials and Methods**

The planet radii are calculated as: $R=M^{1/3.7}$ for rocky cores (*56*), and $R=f\times M^{1/3.7}$ for icy cores (which also contain rock+metal), where f is an increasing function of ice mass fraction "x". For instance, if x=0 (no ice), then f=1; if x=1 (100% ice), f=1.41. Interpolation gives f=(1+0.55 x-0.14 x^2) (*57*). From disk element abundance, "x" of icy cores typically range from 1/2 to 2/3, depending on the incorporation of more volatile clathrate and hydrate ices. Then, "x" is assumed to follow a uniform distribution in between 1/2 (solar rock:$H_2O$-ice ratio) and 2/3 (solar rock:total ices ratio) for icy cores. The masses of rocky and icy cores are assumed to follow their observed mass distributions respectively (histograms on the X-axis in **Fig. 2**). Considering the mass balance between the rock+metal and ices, the number ratio of rocky to icy cores is approximately 1:2, matching the occurrence rate of super-Earths (1-2 $R_\oplus$) versus sub-Neptunes (2-4 $R_\oplus$) (*12*). With geometric transit probability, their observed number by *Kepler* is about equal. Additional ±7% 1-σ Gaussian errors in $R_p$ and ±30% 1-σ Gaussian errors in $M_p$ are included in the simulation to account for the observational uncertainties. See more details in (**SI Appendix**).

**Acknowledgements**: We thank James Kirk, Jane Huang, Alessandro Morbidelli for discussions. This research was partly supported by a grant from the Simons Foundation (SCOL [award #337090] to L.Z.), the Harvard Faculty of Arts and Sciences Dean's Competitive Fund for Promising Scholarship, and the Sandia Z Fundamental Science Program by the Department of Energy National Nuclear Security Administration under Award Numbers DE-NA0001804 and DE-NA0002937 to S. B. Jacobsen (PI) with Harvard University. This research is the authors' views and not those of the DOE. Sandia National Laboratories is a multimission laboratory managed and operated by National Technology and Engineering Solutions of Sandia, LLC., a wholly owned subsidiary of Honeywell International, Inc., for the U.S. Department of Energy's National Nuclear Security Administration under contract DE-NA-0003525.

**References**

1. R. L. Akeson *et al.*, The NASA Exoplanet Archive: Data and Tools for Exoplanet Research. *Publ. Astron. Soc. Pacific*. **125**, 989 (2013).
2. S. E. Thompson *et al.*, Planetary Candidates Observed by Kepler. VIII. A Fully Automated Catalog With Measured Completeness and Reliability Based on Data Release 25. *Astrophys. J. Suppl. Ser. Vol. 235, Issue 2, Artic. id. 38, 49 pp. (2018).* **235** (2017), doi:10.3847/1538-4365/aab4f9.
3. L. Zeng *et al.*, Planet size distribution from the Kepler mission and its implications for planet formation. *Lunar Planet. Sci. Conf.* (2017), p. Abstract #1576, (available at http://adsabs.harvard.edu/abs/2017LPI....48.1576Z).
4. B. J. Fulton *et al.*, The California- Kepler Survey. III. A Gap in the Radius Distribution of Small Planets. *Astron. J.* **154**, 109 (2017).
5. V. Van Eylen *et al.*, An asteroseismic view of the radius valley: stripped cores, not born rocky (2017) (available at http://arxiv.org/abs/1710.05398).
6. A. W. Mayo *et al.*, 275 Candidates and 149 Validated Planets Orbiting Bright Stars in K2 Campaigns 0-10. *Astron. Journal, Vol. 155, Issue 3, Artic. id. 136, 25 pp. (2018).* **155** (2018), doi:10.3847/1538-3881/aaadff.
7. T. A. Berger, D. Huber, E. Gaidos, J. L. van Saders, Revised Radii of Kepler Stars and Planets using Gaia Data Release 2 (2018) (available at http://arxiv.org/abs/1805.00231).
8. B. J. Fulton, E. A. Petigura, The California Kepler Survey VII. Precise Planet Radii Leveraging Gaia DR2 Reveal the Stellar Mass Dependence of the Planet Radius Gap (2018) (available at https://arxiv.org/abs/1805.01453).
9. G. Gaia Collaboration *et al.*, Gaia Data Release 2. Summary of the contents and survey properties (2018) (available at http://arxiv.org/abs/1804.09365).
10. L. Lindegren *et al.*, Gaia Data Release 2: The astrometric solution (2018) (available at http://arxiv.org/abs/1804.09366).
11. K. G. Stassun, E. Corsaro, J. A. Pepper, B. S. Gaudi, Empirical Accurate Masses and Radii of Single Stars with *TESS* and *Gaia*. *Astron. J.* **155**, 22 (2017).
12. E. A. Petigura *et al.*, The California- *Kepler* Survey. IV. Metal-rich Stars Host a Greater Diversity of Planets. *Astron. J.* **155**, 89 (2018).
13. J. E. Owen, Y. Wu, The Evaporation Valley in the Kepler Planets. *ApJ*. **847**, 29 (2017).
14. L. Zeng, S. B. Jacobsen, D. D. Sasselov, A. Vanderburg, Survival Function Analysis of Planet Size Distribution with Gaia DR2 Updates. *MNRAS Accept.* (2018).
15. E. R. Adams, S. Seager, L. Elkins- Tanton, Ocean Planet or Thick Atmosphere: On the Mass- Radius Relationship for Solid Exoplanets with Massive Atmospheres. *Astrophys. J.*

**673**, 1160–1164 (2008).
16. K. I. Öberg, R. Murray-Clay, E. A. Bergin, The effects of snowlines on c/o in planetary atmospheres. *Astrophys. J.* **743**, L16 (2011).
17. C. Hayashi, Structure of the Solar Nebula, Growth and Decay of Magnetic Fields and Effects of Magnetic and Turbulent Viscosities on the Nebula. *Prog. Theor. Phys. Suppl.* **70**, 35–53 (1981).
18. D. J. Stevenson, J. I. Lunine, Rapid formation of Jupiter by diffusive redistribution of water vapor in the solar nebula. *Icarus*. **75**, 146–155 (1988).
19. S. J. Desch, Mass Distribution and Planet Formation in the Solar Nebula. *Astrophys. J.* **671**, 878–893 (2007).
20. K. Zhang, K. M. Pontoppidan, C. Salyk, G. A. Blake, Evidence for a snow line beyond the transitional radius in the TW Hya protoplanetary disk. *Astrophys. J.* **766**, 82 (2013).
21. K. Zhang, G. A. Blake, E. A. Bergin, Evidence of fast pebble growth near condensation fronts in the HL Tau protoplanetary disk. *Astrophys. J.* **806**, L7 (2015).
22. L. A. Cieza *et al.*, Imaging the water snow-line during a protostellar outburst. *Nature*. **535**, 258–261 (2016).
23. J.-E. Lee *et al.*, The ice composition in the disk around V883 Ori revealed by its stellar outburst. *Nat. Astron.*, 1 (2019).
24. A. Morbidelli, Accretion Processes (2018) (available at http://arxiv.org/abs/1803.06708).
25. A. Morbidelli, M. Lambrechts, S. Jacobson, B. Bitsch, The great dichotomy of the Solar System: Small terrestrial embryos and massive giant planet cores. *Icarus*. **258**, 418–429 (2015).
26. A.-M. A. Piso, K. I. Öberg, T. Birnstiel, R. A. Murray-Clay, C/O and snowline locations in protoplanetary disks: the effect of radial drift and viscous gas accretion. *Astrophys. J.* **815**, 109 (2015).
27. D. D. Sasselov, M. Lecar, On the Snow Line in Dusty Protoplanetary Disks. *Astrophys. J.* **528**, 995 (2000).
28. T. Sato, S. Okuzumi, S. Ida, On the water delivery to terrestrial embryos by ice pebble accretion. *A&A*. **589** (2016), doi:10.1051/0004-6361/201527069.
29. R. G. Martin, M. Livio, On the evolution of the snow line in protoplanetary discs. *Mon. Not. R. Astron. Soc. Lett.* **425**, L6–L9 (2012).
30. W. Kley, R. P. Nelson, Planet-Disk Interaction and Orbital Evolution. *Annu. Rev. Astron. Astrophys.* **50**, 211–249 (2012).
31. S. N. Raymond *et al.*, Planet-planet scattering leads to tightly packed planetary systems. *Astrophys. J.* **696**, L98–L101 (2009).
32. A. Johansen, M. Lambrechts, Forming Planets via Pebble Accretion. *Annu. Rev. Earth Planet. Sci.* **45**, 359–387 (2017).
33. J. A. Kegerreis *et al.*, Consequences of Giant Impacts on Early Uranus for Rotation, Internal Structure, Debris, and Atmospheric Erosion. *Astrophys. J.* **861**, 52 (2018).
34. R. R. Rafikov, Atmospheres of Protoplanetary Cores: Critical Mass for Nucleated Instability. *Astrophys. J.* **648**, 666–682 (2006).
35. N. K. Inamdar, H. E. Schlichting, The formation of super-Earths and mini-Neptunes with giant impacts. *Mon. Not. R. Astron. Soc.* **448**, 1751–1760 (2015).
36. R. A. Marcus, S. T. Stewart, D. Sasselov, L. Hernquist, Collisional stripping and disruption of super-earths. *Astrophys. J.* **700**, L118–L122 (2009).
37. J. M. Brewer, S. Wang, D. A. Fischer, D. Foreman-Mackey, Compact Multi-planet

Systems are more Common around Metal-poor Hosts. *Astrophys. J.* **867**, L3 (2018).
38. A. Levi, D. Sasselov, M. Podolak, Volatile Transport inside Super-Earths by Entrapment in the Water-ice Matrix. *Astrophys. J.* **769**, 29 (2013).
39. A. Levi, D. Sasselov, M. Podolak, Structure and Dynamics of Cold Water Super-Earths: The Case of Occluded CH4 and its Outgassing. *Astrophys. Journal, Vol. 792, Issue 2, Artic. id. 125, 44 pp. (2014).* **792** (2014), doi:10.1088/0004-637X/792/2/125.
40. A. Levi, S. J. Kenyon, M. Podolak, D. Prialnik, H-atmospheres of Icy Super-Earths Formed In Situ in the Outer Solar System: An Application to a Possible Planet Nine. *Astrophys. J.* **839**, 111 (2017).
41. F. Soubiran, B. Militzer, Miscibility calculations for water and hydrogen in giant planets. *Astrophys. J.* **806**, 228 (2015).
42. R. Hu, S. Seager, Y. L. Yung, Helium atmospheres on warm neptune- and sub-neptune-sized exoplanets and applications to GJ 436b. *Astrophys. J.* **807**, 8 (2015).
43. D. R. Anderson *et al.*, The discoveries of WASP-91b, WASP-105b and WASP-107b: Two warm Jupiters and a planet in the transition region between ice giants and gas giants. *Astron. Astrophys.* **604**, A110 (2017).
44. J. J. Spake *et al.*, Helium in the eroding atmosphere of an exoplanet. *Nature*. **557**, 68–70 (2018).
45. L. Nortmann *et al.*, Ground-based detection of an extended helium atmosphere in the Saturn-mass exoplanet WASP-69b (2018), doi:10.1126/science.aat5348.
46. M. Salz *et al.*, Detection of He I $\lambda10830$ \AA{} absorption on HD 189733 b with CARMENES high-resolution transmission spectroscopy (2018), doi:10.1051/0004-6361/201833694.
47. R. Allart *et al.*, Spectrally resolved helium absorption from the extended atmosphere of a warm Neptune-mass exoplanet (2018), doi:10.1126/science.aat5879.
48. M. Mansfield *et al.*, Detection of Helium in the Atmosphere of the Exo-Neptune HAT-P-11b (2018), doi:10.3847/2041-8213/aaf166.
49. M. Millot *et al.*, Experimental evidence for superionic water ice using shock compression. *Nat. Phys.*, 1 (2018).
50. J. Farihi, B. T. Gänsicke, D. Koester, Evidence for Water in the Rocky Debris of a Disrupted Extrasolar Minor Planet. *Science (80-. ).* **342** (2013) (available at http://science.sciencemag.org/content/342/6155/218).
51. R. Raddi *et al.*, Likely detection of water-rich asteroid debris in a metal-polluted white dwarf (2015), doi:10.1093/mnras/stv701.
52. H. R. Wakeford *et al.*, The Complete transmission spectrum of WASP-39b with a precise water constraint (2017), doi:10.3847/1538-3881/aa9e4e.
53. J. M. Brewer, D. A. Fischer, C/O and Mg/Si Ratios of Stars in the Solar Neighborhood. *Astrophys. J.* **831**, 20 (2016).
54. A. Bhattacharya *et al.*, WFIRST Exoplanet Mass Measurement Method Finds a Planetary Mass of $39\pm 8 M_\oplus$ for OGLE-2012-BLG-0950Lb (2018) (available at https://arxiv.org/abs/1809.02654).
55. D. Suzuki *et al.*, The exoplanet mass-ratio function from the MOA-II survey: discovery of a break and likely peak at a Neptune mass. *Astrophys. J.* **833**, 145 (2016).
56. L. Zeng, D. D. Sasselov, S. B. Jacobsen, Mass-Radius Relation for Rocky Planets Based on PREM. *Astrophys. J.* **819**, 127 (2016).
57. L. Zeng, D. Sasselov, A Detailed Model Grid for Solid Planets from 0.1 through 100 Earth

Masses. *Publ. Astron. Soc. Pacific*. **125**, 227–239 (2013).
58. M. I. Petaev, The GRAINS thermodynamic and kinetic code for modeling nebular condensation. *Calphad Comput. Coupling Phase Diagrams Thermochem.* **33**, 317–327 (2009).
59. J. S. Lewis, J. S., Low temperature condensation from the solar nebula. *Icarus*. **16**, 241–252 (1972).
60. J. W. Chamberlain, D. M. Hunten, *Theory of planetary atmospheres : an introduction to their physics and chemistry* (Academic Press, 1987).

**Figures:**

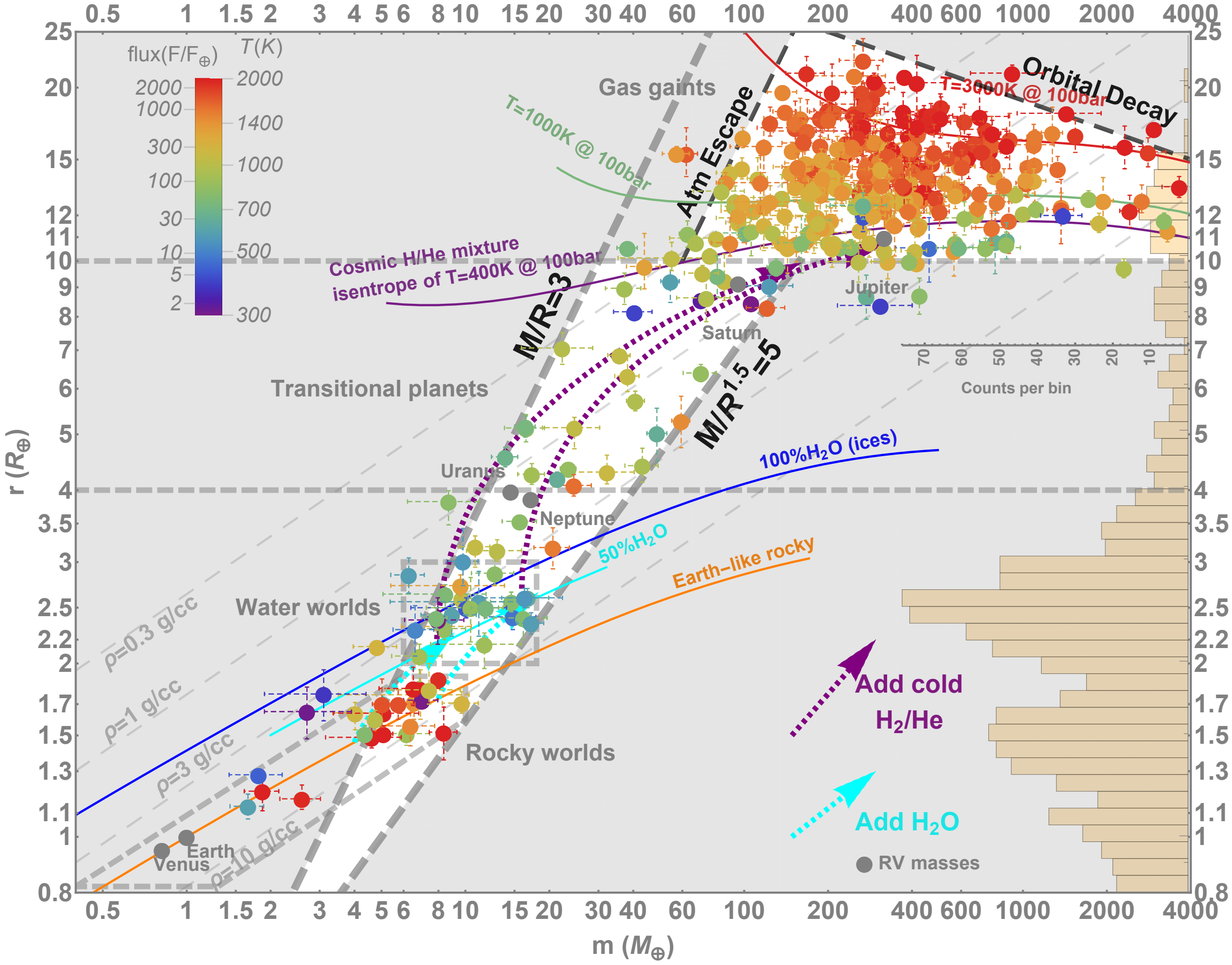

**Fig. 1** The Mass-Radius variations among selected exoplanets with masses determined by the radial-velocity (RV) method and densities constrained to better than ±50% (1-σ). The plotted data are listed in (**SI Appendix, Table S1**). The color of the data points denotes stellar insolation (see legend in the upper left corner) in the Earth units (expressed as either the amount of stellar bolometric radiation reaching a given area at their orbital distances, assuming negligible orbital eccentricities, normalized to the Earth's value or surface equilibrium temperatures assuming Earth-like albedo). The vertical histogram on the right Y-axis shows the log-binned radius distribution of 1156 *Kepler* confirmed/candidate planets with radius errors less than ±10% (1-σ, the average error is about ±7%), orbiting only the main-sequence host stars within the effective temperatures in the 5000-6500 K range (Gaia Data Release 2 (*7*)). The dotted cyan and purple arrows show the growth trajectories of planets formed by continuous addition of either $H_2O$-ices or $H_2$+He gas to a planetary core of a given mass (**SI Appendix**). The area outlined by the gray dashed rectangle is shown in **Fig. 2**.

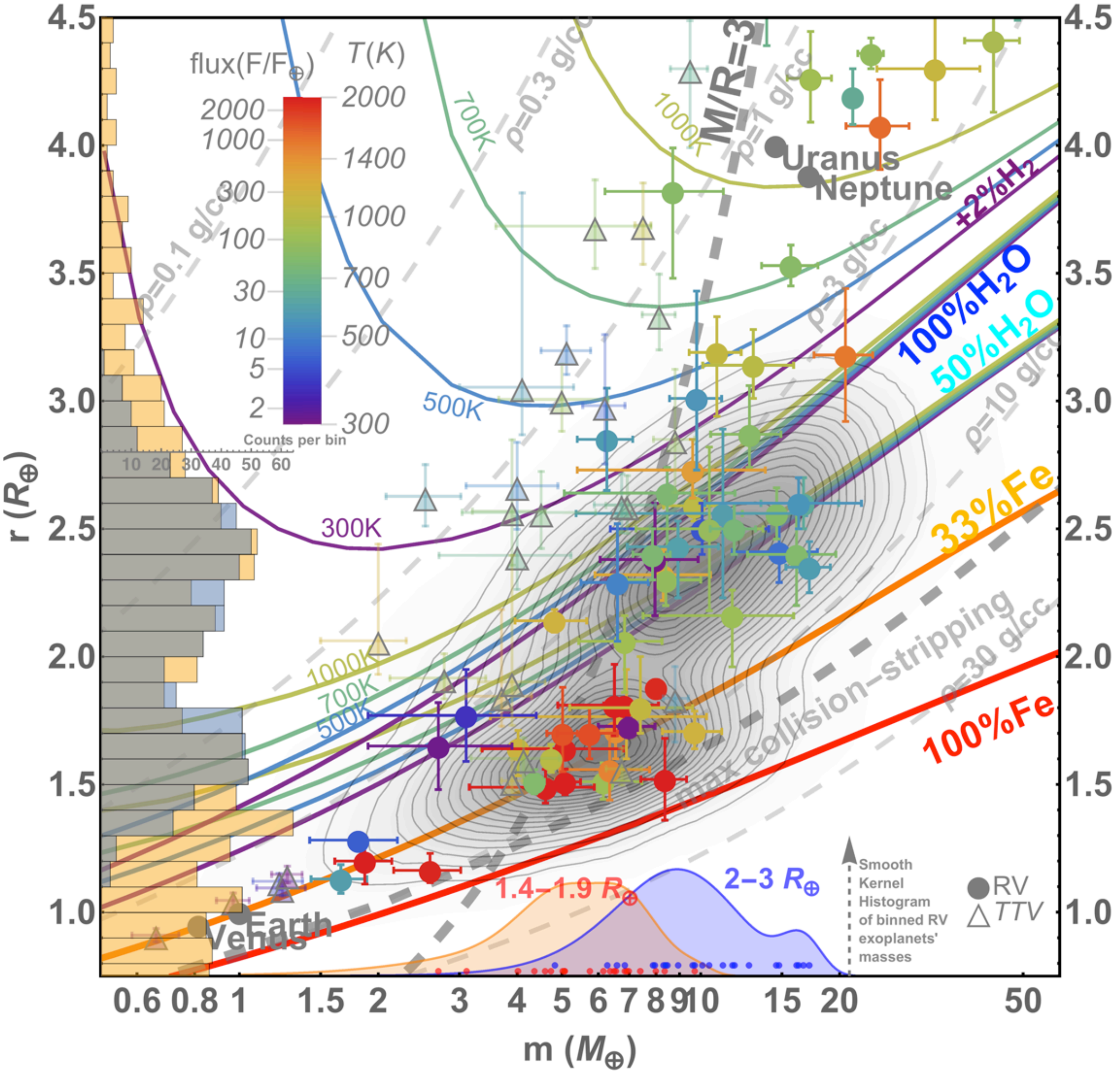

**Fig. 2** Blowup of **Fig. 1**. Radius gap at 2 R⊕ separates two distinctive groups of RV planets (1.4-1.9 R⊕ and 2-3 R⊕). Their smooth kernel mass distributions on the bottom X-axis show a significant offset, with truncation of the super-Earths (yellow) and sub-Neptunes (purple) at ~10 M⊕ and ~20 M⊕, respectively. The histogram on the left Y-axis compares the results of Monte-Carlo simulation (light blue) with the observations (yellow). Two sets of $H_2O$ M-R curves (blue - 100 mass% $H_2O$, cyan - 50 mass% $H_2O$; cores consist of rock and $H_2O$ ice in 1:1 proportion by mass) are calculated for an isothermal fluid/steam envelope at 300K, 500 K, 700 K, 1000 K, sitting on top of ice VII-layer at the appropriate melting pressure. A set of mass-radius curves (upper portion of the diagram) is calculated for the same temperatures assuming the addition of an isothermal 2 mass % $H_2$-envelope to the top of the 50 mass% $H_2O$-rich cores.

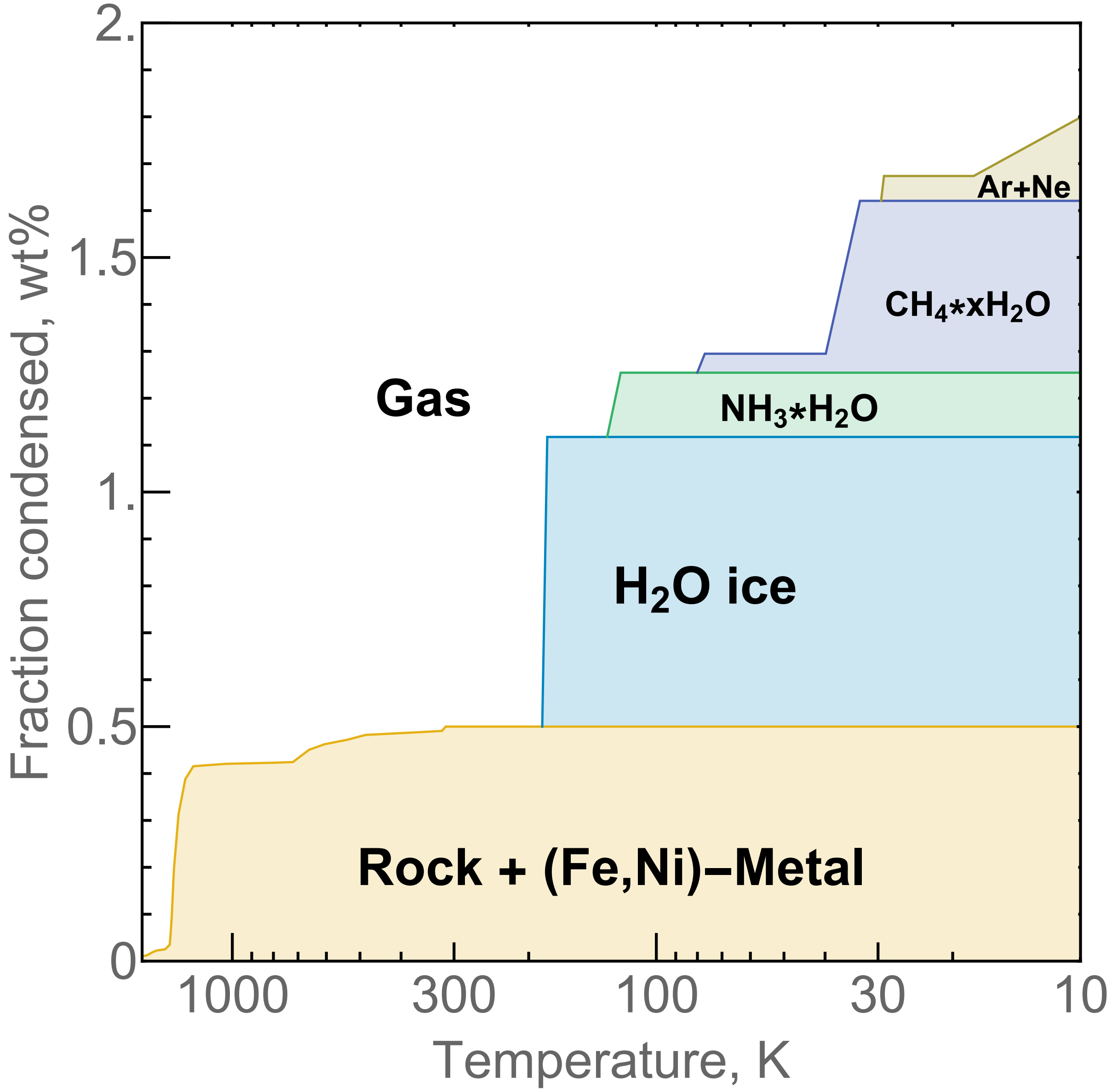

**Fig. 3** The stability fields of planet-building materials in a protoplanetary disk of solar composition at $10^{-4}$ bar. The fraction of condensed rocky matter (>200 K) is calculated with the GRAINS code (*58*) while condensation temperatures and abundances of ices are taken from (*59*). The composition of gas is temperature-dependent and always complementary to the composition of the condensed phase assemblage.

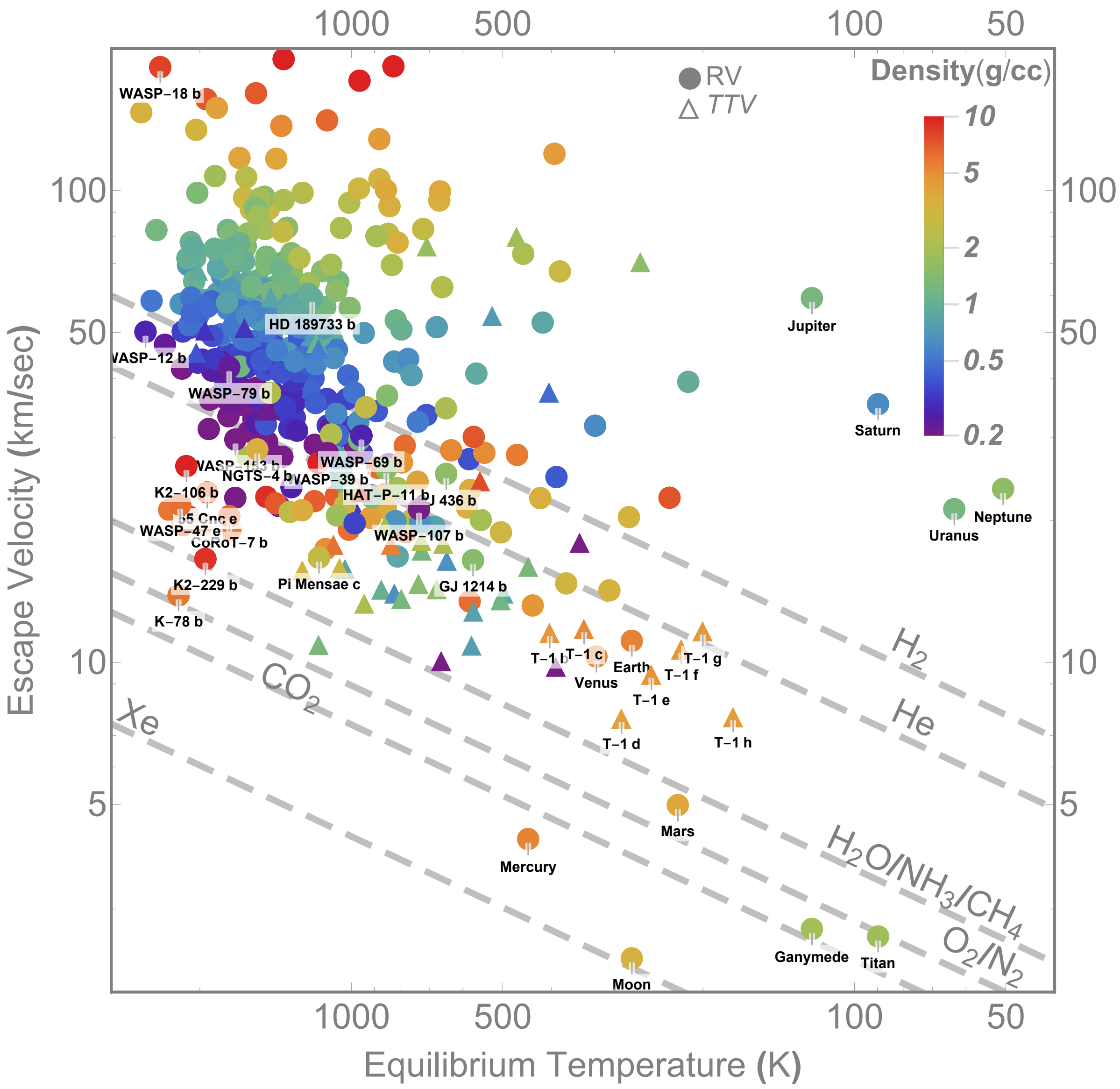

**Fig. 4** Atmospheric escape velocities versus surface equilibrium temperature. Data from (*60*). Color coding is the bulk density of planet. Many super-Earths and sub-Neptunes are not expected to retain a $H_2$-He-dominated envelope over a billion-year timescale.

Supplementary Information for

# Growth Model Interpretation of Planet Size Distribution

**Authors:** Li Zeng*[1,2], Stein B. Jacobsen[1], Dimitar D. Sasselov[2], Michail I. Petaev[1,2], Andrew Vanderburg[3], Mercedes Lopez-Morales[2], Juan Perez-Mercader[1], Thomas R. Mattsson[4], Gongjie Li[5], Matthew Z. Heising[2], Aldo S. Bonomo[6], Mario Damasso[6], Travis A. Berger[7], Hao Cao[1], Amit Levi[2], Robin D. Wordsworth[1].

*Correspondence to: astrozeng@gmail.com

**This PDF file includes:**

Materials and Methods
Figure S1
Figure S2
Table S1
References for SI reference citations

## Materials and Methods

**Planet data** of all exoplanets used for our analysis have been downloaded from the *NASA Exoplanet Archive* (*1*, *2*) on several occasions, with the manual addition/modification of some newly-discovered planets: in particular, on Nov.10, 2016 for initial mass-radius plot and statistical assessments, and subsequently on Feb.10, May 31, Sept.9, Oct.16, 2017, and Feb.20, Jun.15, 2018 for updates. Moreover, we have updated the radii of some planets below 8 $R_\oplus$ with GAIA DR1 (*3*) and GAIA DR2 (*4*). To make the mass-radius plot more precise, we select planets with density constraints better than about ±50%.
Specifically, since density $\rho \propto M/R^3$, taking natural logarithmic gives $\ln\rho = \ln M - 3\ln R + const$, and differentiation of it gives: $\frac{d\rho}{\rho} = \frac{dM}{M} - 3 \cdot \frac{dR}{R}$.

Error propagation assumes errors of mass and radius are independent: $\frac{\sigma\rho}{\rho} \approx \sqrt{\left(\frac{\sigma M}{M}\right)^2 + \left(3 \cdot \frac{\sigma R}{R}\right)^2}$
Then, taking into account asymmetric error bars to avoid planets with an extremely large error bar in one direction, we chose the selection criterion as:

$$\ln\left(\frac{\rho + \sigma\rho^+}{\rho - \sigma\rho^-}\right) \approx \sqrt{\left[\ln\left(\frac{M + \sigma M^+}{M - \sigma M^-}\right)\right]^2 + \left[3 \cdot \ln\left(\frac{R + \sigma R^+}{R - \sigma R^-}\right)\right]^2} \lesssim 1$$

where the + and – signs denote the error bar in the plus and minus direction respectively. The planet data are listed in (**SI Appendix, Table S1**), with detailed references (*3–80*).

**Figs. 1, 2, & 4** have been plotted using this selection criterion with the built-in schemes in *Mathematica (Version 11)* including "ListLogLogPlot" and "ErrorListPlot".

**Statistical analysis** of the planet data has been performed with *Mathematica*'s built-in "WeightedData", "SmoothHistogram", and "SmoothKernelDistribution" packages. The smoothing is done with the *Mathematica*'s default "Silverman" bandwidth-selection-method. Each planet data is weighted by the product of its positive and negative 1-σ mass error bars. We then calculate the weighted mean and weighted standard deviation for each set of planets.

**Observations bias** of the RV method tends to detect and characterize planets of higher masses and orbiting closer to their host stars (*81*). Because of this bias, the RV method becomes more sensitive as mass of a planet increases at given orbit, with current detection limit of ~3 $M_\oplus$. When we invert the RV mass distribution to a radius distribution, we miss the planets smaller than ~1.4 $R_\oplus$. Because such planets are not taken into account by our simulations, our Monte Carlo histogram (gray bars in **Fig. 2**) has much less planets below ~1.4 $R_\oplus$ than observed by the transiting method. This also results in a slightly higher radius gap compared to the population detected by the transiting method. Since this paper considers planets more massive than ~3 $M_\oplus$, the observational bias of the RV method has little effect on our results and conclusions.

**Multi-planet systems** which contain both super-Earths and sub-Neptunes are interesting. First of all, the goal of our paper is not to model a specific exoplanetary system but rather to use the observed statistical distributions for identifying possible nature of different planet populations. This is why we focus on the majority of exoplanets found by the *Kepler* mission. Our growth model is statistical in nature and does not intend to explain detailed architecture of a specific planetary system. Such outliers and extreme cases constitute ~5% of the observed exoplanet population which our model cannot explain. Indeed, real planet-forming scenarios are more complicated and, perhaps, somewhat different around different stars. One of the critical features in different disks would be how and when the snowlines and planet-building materials migrate during planet formation. For example, taking into account formation of a planet at a snowline followed by its inward migration, Ormel et al. 2017 (*82*) managed to explain the formation and architecture of compact multi-planet systems such as *Trappist*-1 (*70*, *76*, *83*). *Kepler*-20 (*26*, *84*, *85*) can be also explained by such scenario. We call this scenario the "snowline conveyor belt hypothesis" which is consistent with our growth model. In addition, planet-planet collisions (i.e. giant impacts), and more generally, strong dynamical interactions among planets during planet formation have long been suggested by planet formation theory, and recently have been confirmed by the observation and characterization of the *Kepler*-107 system (*86*, *87*).

**TTV planets**. The mass determination from transit-timing-variation (TTV) for multi-planet systems is known to be difficult and, often requires a later revision, because the masses determined are often degenerate with the combined eccentricities of planets in that system (*88*). Sometimes the N-body Markov-Chain-Monte-Carlo (MCMC) approach is used to find the best solutions. When a longer time duration of TTV observations becomes available, or data are re-assessed by a different group, the planet masses can change significantly; e.g., *Trappist*-1 system (*70*, *83*, *89*, *90*) and *Kepler*-11 system (*68*, *71*, *72*). Some TTV planets (**Fig. 2**) of 2-4 $R_\oplus$ seem to be puffy, i.e., of low-bulk-density. They can be explained by a core (rocky/icy, 3-10 $M_\oplus$) plus a thick, but very light (1~2% of total planet mass) gaseous envelope. Also, statistically speaking, systems with significant TTV effects are less common among all *Kepler* planetary systems; e.g., among 2599 *Kepler* Objects of Interest (KOIs) only 260 were found with significant TTVs with long-term (>100 day) variations (*67*). TTV systems are generally characterized by tight spacing of planets, circular (low ***e***) and coplanar (low ***i***) orbits, and low obliquities of the host stars (*91*). The difference in planet masses between TTV and RV may be correlated with the host stellar metallicities – TTV planets prefer low-metallicity host stars (*92*). They can be explained by planet formation under different disk conditions, e.g., solid surface density versus gas surface density in the disk (*93*, *94*).

**Transitional planets** (of 4-10 $R_\oplus$), in particular the ones slightly larger than 4 $R_\oplus$ residing to the right of the purple growth curves in **Fig. 1**, have massive cores ($\gtrsim$20 $M_\oplus$) but not much gas ($\lesssim$20% by mass). For example, K2-55 b (stellar [Fe/H]=0.376±0.095) has core mass estimated to be ~40 $M_\oplus$ with about 10% $H_2$/He envelope by mass, regardless of the core composition (rocky or icy) (*95*). Because their core masses are apparently above the critical core mass (*96–98*), it is interesting to understand how they could have formed with that much solid material without experiencing a runaway gas accretion. One hypothesis is that they arise as mergers of two or more less massive cores. The merging processes tend to merge the denser cores together but (partially) erode their gas envelopes (*99*, *100*), resulting in a net mass increase accompanied by a very small radius increase, that moves them to the right on the mass-radius diagram. Disks with high metallicity may facilitate mergers. The second hypothesis is motivated by the results of micro-lensing surveys (*101*, *102*), which are sensitive to distant planet populations orbiting their host stars typically at a few a.u. The results show that planets in the 20-80 $M_\oplus$ range at a few a.u. are abundant, which is contradictory to the paucity of planets of this mass range predicted by runaway gas accretion models. This suggests that the current runaway core accretion models may require significant revisions to slow down the growth of planets of 10-50 $M_\oplus$ and start the runaway accretion at higher masses of 50-80 $M_\oplus$ instead of 10 $M_\oplus$. The third hypothesis assumes that part of planet formation takes place when the gas in the disk is depleted; cf. theoretical calculations of transitional disks (*103*) and the inferred rapid gas depletion by a few million years from the ALMA survey of the Lupus Proto-planetary disks (*104*, *105*). However, one has to be cautious that the distinction between the total gas depletion and CO depletion cannot be inferred from observation directly, because the inferred $H_2$ depletion is calculated based on the observed CO depletion and assumed ISM-like [CO]/[$H_2$] ratio (per. comm. Jane Huang and Sean

Andrews). The cores of these transitional planets, regardless of a hypothesis of their origin, likely contain large fractions of ices in addition to rocks, just as Uranus and Neptune.

**Mass-radius curves.** The mass-radius curves (**Figs. 1 & 2**) allow calculations of average densities and further constraining the planet bulk compositions and internal structures following the methods that we published earlier (*106–109*) and similar efforts from other researchers in our field (*110–120*). The $H_2$-He is assumed to be a cosmic mixture of 75% $H_2$ and 25% He by mass (*121–123*). The Earth-like composition is assumed to be 32.5 wt.% Fe/Ni-metal plus 67.5 wt.% $MgSiO_3$-rock. The bulk $H_2O$ in deep interior is assumed to be in solid phase along the melting curve (liquid-solid phase boundary) (*124–131*). The colored $H_2O$ M-R curves correspond to isothermal vapor/liquid/super-critical fluid envelope at various surface temperatures: 300K, 500 K, 700 K, and 1000 K, on top of ice VII-layer along the appropriate melting pressure (*132–134*). Since ices all have similar densities, and the $H_2O$-$NH_3$-$CH_4$ mixture is always dominated by $H_2O$ due to chemistry evaluation, the mass-radius curve of the $H_2O$-$NH_3$-$CH_4$ mixture is expected to be very similar to that of pure-$H_2O$. On the other hand, the $H_2$-He mass-radius curves are calculated along various interior adiabats at different internal specific entropies, labelled by the temperature of the corresponding specific entropy at 100-bar level in the gas envelope (400K, 1000K, 3000K). The mass-radius curves are almost parallel to one another and do not criss-cross, as in all cases at high pressure the material structure is supported mainly by the electron degeneracy pressure, which has similar functional dependence on compression. Therefore, the density of a particular material under high pressure is primarily determined by its average atomic weight, which is very different for the three major planet-building materials: rocks, ices, $H_2$-He gas, resulting in significant density differences among them in planetary deep interiors, and secondarily by the crystal structure or temperature. The surface of a planet in our calculation is always defined at 10^-3 bar (1 milli-bar) level as an approximate for the level that transiting observations probe.

**$H_2O$ EOS.** The **equations of states (EOS)** of ice and ice-mixtures under appropriate pressure and temperature conditions are required to understand the interiors of these water worlds. Density Functional Theory calculations based on quantum mechanics have provided thermo-physical properties of water at extreme conditions, including predictions of superionic phases (*127*, *129*). Importantly, key results from DFT simulations have been validated by experiments on Sandia's Z-machine (*135*) as well as by direct measurements of proton conductivity in diamond anvil cell experiments (*136*) and shock compression experiments (*128*). Various models of planet interiors utilizing those EOSs have been introduced (*137*). These models make predictions of the existence of intriguing phases such as the super-ionic state of hot ices and also the non-dipole magnetic field generated in the interior of such planets. They also show that $H_2O$ EOS commonly used in some previous planetary models significantly overestimate the compressibility, and consequently overestimate the density of water at a few hundred GPa (*135*), relevant to the interior of these intermediate-sized planets. This new water EOS (*135*) is adopted for our calculation.

**Growth curves.** The growth curves of either adding $H_2O$-ice (cyan) or $H_2$-He gas (purple) to a core are calculated (**Figs. 1 & 2**). The cyan growth curves start with 4 and 8 $M_\oplus$ rocky cores and show the trajectory of a growing planet on the mass-radius diagram with increasing amount of $H_2O$ added. The purple growth curves start with 8 and 16 $M_\oplus$ icy cores (half-ice and half-rock by mass) and show the trajectory of a growing planet on the mass-radius diagram with increasing amount of cold $H_2$-He gas added. The tangential slope of the growth curve on the mass-radius diagram depends on the ratio between the density of the added material versus the density of the already-existing material on the planet. The initial slope of a growth curve on a log-log mass-radius plot can be approximated as $d\ln R/d\ln M \sim \rho_{core}/(3\rho_{acc})$, where $\rho_{core}$ is the core density and $\rho_{acc}$ is the density of added material. The tangential slope of the growth curve of $H_2$-He is nearly vertical initially, due to the very low density of gas, regardless of the core composition---rocky or icy. The $H_2$-He gaseous envelope of 1-2 $R_\oplus$ thick is insignificant in mass---a few percent in mass% (*138*, *139*). Until more and more $H_2$-He is added so that the $H_2$-He in the planetary deep interiors become self-compressed to high enough densities, so that the tangential slope of $H_2$-He growth curve reduces. A *Mathematica* program for this calculation is available upon request. However, be cautious that a simple growth curve is not to be understood as a realistic time-evolution for a single planet growth, such that a planet core must accrete all rocks before accreting any ices. But rather, it is a statistical trend followed by comparing the overall planet populations. In fact, it is more likely that the earlier condensed dust in the nebula is entrapped into multi-layers of ices---icy pebbles (**SI Appendix, Fig. S1**). Then, a planet core, if formed beyond or near the snowline, or if accreting icy pebbles, would accrete rocks and ices simultaneously. The dichotomy of rocky versus icy bodies is also reflected in the densities of solar system giant planets' satellites (**SI Appendix, Fig. S2**).

**Monte Carlo simulation.** We conduct a series of Monte Carlo simulations, in order to test our model of rocky and intermediate-sized planetary mass-radius relation, which provide clues on the composition of the planets. Specifically, we first sample the planetary masses using the observed mass distribution, where the masses of the rocky worlds are generated following the empirical smooth kernel distribution of planets with radii between 1.4-1.9 $R_\oplus$, and the masses of the icy worlds follow the empirical smooth kernel distribution of those with radii between 2-3 $R_\oplus$. Then, we calculate the radii of the planets based on the mass-radius relation: the radii of the rocky worlds can be obtained equating $R=M^{1/3.7}$ (*108*, *140*), and the radii of the intermediate-sized worlds can be calculated as $R=f\times M^{1/3.7}$. f is an indicator of the composition of the intermediate-sized planet. For instance, f=1, if the intermediate-sized worlds still follow the rocky planetary mass-radius relation. f=(1+0.55 x-0.14 x^2), where the ice mass fraction "x" of the icy world is around 1/2~2/3, depending on incorporating the more volatile (C,N)-clathrate-ices or not. For a complete water world (100% water), f=1.41. In order to compare with the observed exoplanets with RV mass measurements, we assume that "x" follows a uniform distribution in between 1/2 and 2/3 for the icy worlds, corresponding to fully incorporating $H_2O$ ices in proportion to rocks, and partially incorporating (C,N)-clathrate-ices in some cases. Then we include an additional ±7% uncertainty to account for the observational errors in planetary radii, and an additional ±30% uncertainty to account for the observational errors in planetary masses. Combining the Monte Carlo results of these two populations and adopt a number ratio of #rocky cores : #icy cores = 1 : 2, we recover the bimodality with the planet radius gap matches the observation in the mass-radius diagram (**Fig. 2**). The generated 2-dimensional mass-radius histogram and the 1-dimensional radius histogram are only meant to be compared with the observed planet populations in between 1.4-3 $R_\oplus$, since our assumption in generating this Monte Carlo simulation draws the empirical mass distribution only from within this radius range. Therefore, planets larger than about 3 $R_\oplus$ generally require the presence of some gaseous envelope, and this is beyond the scope of this simulation. This gas envelope could either be primordial $H_2$-He-dominated nebular gas or could be formed by the dissociation and vaporization of the ices near the surface and in the atmosphere of these planets. This **Monte Carlo** simulation is performed using *Mathematica*'s built-in random number generator "RandomVariate". "DensityHistogram" is used for the 2-dimensional plot of the probability density contours from the Monte Carlo simulation results. A *Mathematica* program for generating the Monte Carlo simulation is available upon request.

**Supplementary Figures:**

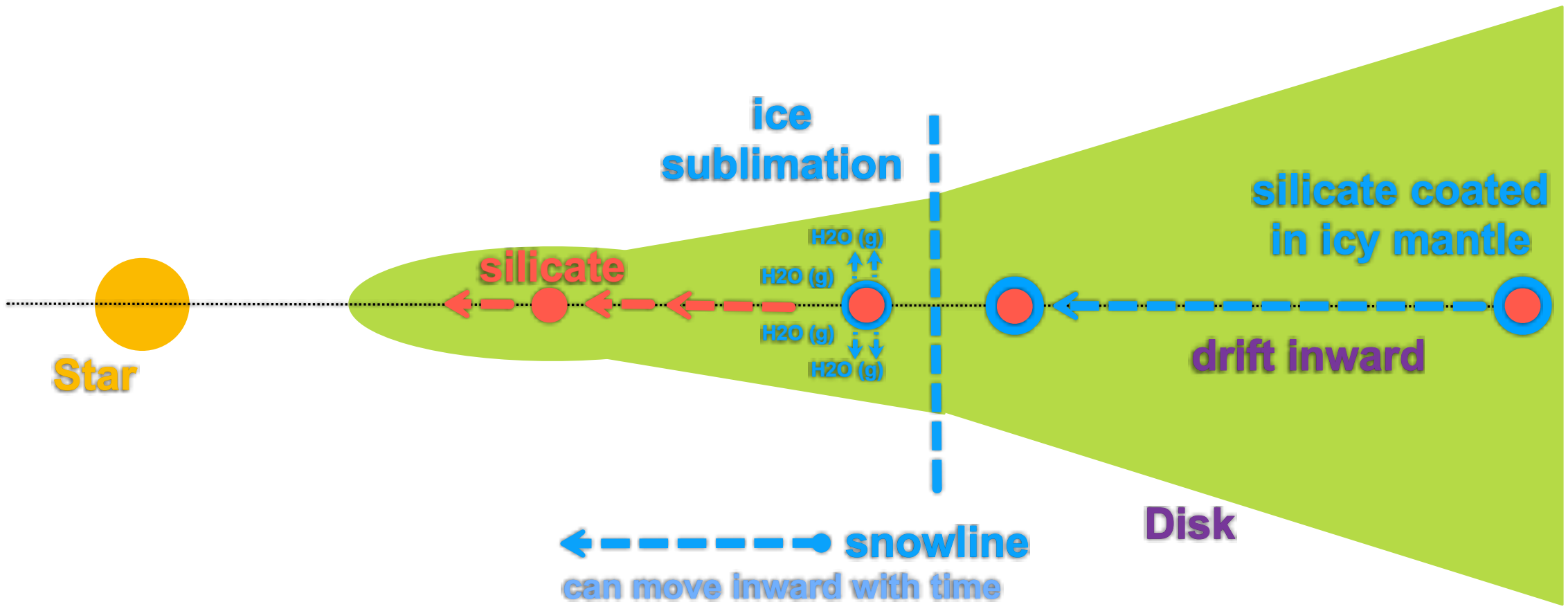

**Fig. S1** Cartoon of a protoplanetary disk showing the inward drift of solid particles consisting of silicates (red) and ices (blue). When particles cross the snowline (vertical blue dashed line) they start losing ice by sublimation. Larger particles of pebble size may drift deeper into the inner zone until they completely loose ice. The snowline itself would move inward as the disk cools with time (*141–143*). Large icy planetary cores formed beyond snowline can later migrate inward due to planet-disk interactions (*144–146*) or planet-planet scattering (*147*).

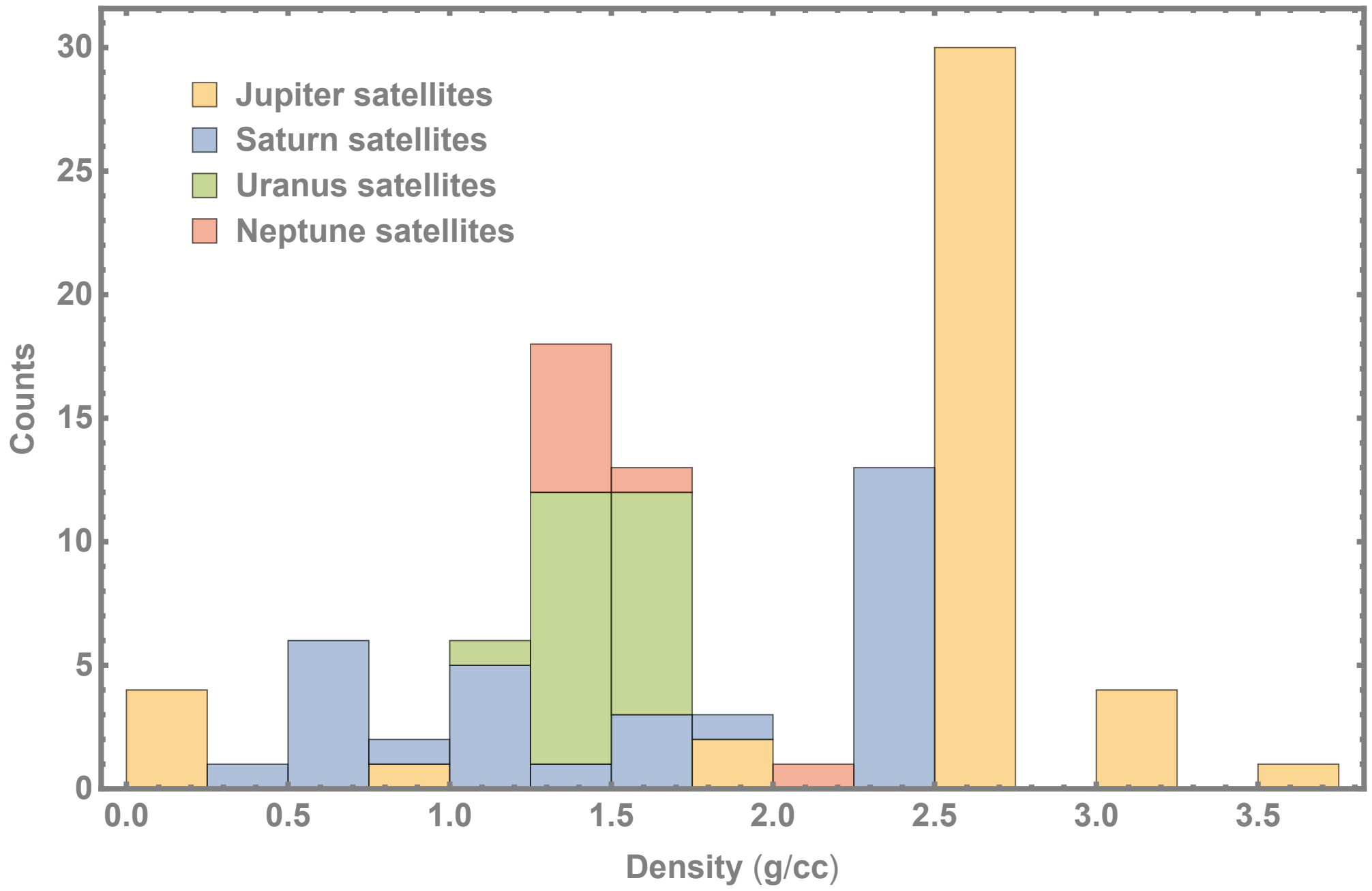

**Fig. S2** Density variations among satellites of the outer Solar System planets. The two well-defined peaks at ~1.5 and ~2.5 g/cc suggest that the planet formation process involved interactions between both icy and rocky bodies.

**Table S1**.

# <u>ref.</u> (*3*–*80*).
# This file is modified from the NASA Exoplanet Archive (*1*, *2*): http://exoplanetarchive.ipac.caltech.edu
# COLUMN pl_orbper: Orbital Period [days]
# COLUMN pl_ttvflag: TTV Flag (1=yes, 0=no)
# COLUMN st_teff: Host Stellar Effective Temperature [K]
# COLUMN st_mass: Host Stellar Mass [Solar mass = 1.9885 * 10^30 kg]
# COLUMN st_rad: Host Stellar Radius [Solar radii = 6.957 * 10^8 meter]
# COLUMN pl_name: Planet Name
# COLUMN pl_rvflag: Planet RV Flag (1=yes, 0=no)
# COLUMN pl_masse: Planet Mass [Earth mass = 5.972 * 10^24 kg]
# COLUMN pl_masseerr1: Planet Mass Upper Unc. [Earth mass]
# COLUMN pl_masseerr2: Planet Mass Lower Unc. [Earth mass]
# COLUMN pl_rade: Planet Radius [Earth radii = 6.371 * 10^6 meter]
# COLUMN pl_radeerr1: Planet Radius Upper Unc. [Earth radii]
# COLUMN pl_radeerr2: Planet Radius Lower Unc. [Earth radii]
# COLUMN met: Host Stellar Metallicity [Fe/H]
# COLUMN mass_ref: Reference to the Planet Mass adopted
# COLUMN radius_ref: Reference to the Planet Radius adopted
# COLUMN additional_ref: Additional Reference
#

| # | pl_orbper | pl_ttvflag | st_teff | st_mass | st_rad | pl_name | pl_rvflag | pl_masse | pl_masseerr1 | pl_masseerr2 | pl_rade | pl_radeerr1 | pl_radeerr2 | met | mass_ref | radius_ref | add_ref |
|---|---|---|---|---|---|---|---|---|---|---|---|---|---|---|---|---|---|
| | | | | | | | | **RV planets < 1.4 Earth radii** | | | | | | | | | |
| 1 | 0.584249 | 0 | 5185 | 0.837 | 0.793 | K2-229 b | 1 | 2.59 | 0.43 | -0.43 | 1.164 | 0.066 | -0.048 | -0.09 | Santerne_2018 | Santerne_2018 | |
| 2 | 1.6289287 | 0 | 3270 | 0.181 | 0.211 | GJ 1132 b | 1 | 1.66 | 0.23 | -0.23 | 1.13 | 0.056 | -0.056 | null | Bonfils_2018 | Dittman_2017 | Diamond-Lowe_2018 |
| 3 | 0.355007 | 0 | 5058 | 0.76 | 0.74 | K-78 b | 1 | 1.87 | 0.27 | -0.26 | 1.2 | 0.09 | -0.09 | -0.14 | Grunblatt_2015 | Grunblatt_2015 | |
| 4 | 3.777931 | 0 | 3216 | 0.179 | 0.2139 | LHS 1140 c | 1 | 1.81 | 0.39 | -0.39 | 1.282 | 0.024 | -0.024 | -0.24 | Ment_2018 | Ment_2018 | Dittmann_2017 |

**RV planets 1.4-2.0 Earth radii**

| | | | | | | | | | | | | | | | | | |
|---|---|---|---|---|---|---|---|---|---|---|---|---|---|---|---|---|---|
| 1 | 2.17 | 0 | 4503 | 0.7 | 0.72 | K2-216 b | 1 | 7.4 | 2.2 | -2.2 | 1.8 | 0.2 | -0.1 | -0.18 | Persson_2018 | Persson_2018 | Mayo_2018 |
| 2 | 0.7365 | 0 | 5172 | 0.91 | 0.94 | 55 Cnc e | 1 | 8 | 0.3 | -0.3 | 1.88 | 0.03 | -0.03 | 0.31 | Demory_2016 | Bourrier_2018 | Winn_2011 |
| 3 | 0.8536 | 0 | 5259 | 0.91 | 0.82 | CoRoT-7 b | 1 | 5.74 | 0.86 | -0.86 | 1.7 | 0.1 | -0.1 | 0.05 | Barros_2014 | Stassun_2017 | Haywood_2014 |
| 4 | 3.0929 | 0 | 4699 | 0.81 | 0.78 | HD 219134 b | 1 | 4.74 | 0.19 | -0.19 | 1.602 | 0.055 | -0.055 | 0.11 | Gillon_2017 | Gillon_2017 | Motalebi_2015 |
| 5 | 6.7646 | 0 | 4699 | 0.81 | 0.78 | HD 219134 c | 1 | 4.36 | 0.22 | -0.22 | 1.511 | 0.047 | -0.047 | 0.11 | Gillon_2017 | Gillon_2017 | Motalebi_2015 |
| 6 | 0.9596 | 0 | 5261 | 0.86 | 0.86 | HD 3167 b | 1 | 5.02 | 0.38 | -0.38 | 1.7 | 0.18 | -0.15 | 0.04 | Christiansen_2017 | Christiansen_2017 | Gandolfi_2017 |
| 7 | 24.64 | 0 | 3896 | 0.6 | 0.56 | K2-3 c | 1 | 3.1 | 1.3 | -1.2 | 1.77 | 0.18 | -0.18 | -0.01 | Damasso_2018 | Damasso_2018 | Sinukoff_2016 |
| 8 | 44.56 | 0 | 3896 | 0.6 | 0.56 | K2-3 d | 1 | 2.7 | 1.2 | -0.8 | 1.65 | 0.17 | -0.17 | -0.01 | Damasso_2018 | Damasso_2018 | Sinukoff_2016 |
| 9 | 0.5713 | 0 | 5470 | 0.945 | 0.87 | K2-106 b | 1 | 8.36 | 0.96 | -0.94 | 1.52 | 0.16 | -0.16 | -0.025 | Guenther_2017 | Guenther_2017 | Sinukoff_2017 Adams_2017 |
| 10 | 0.3693 | 0 | 5200 | 0.84 | 0.81 | K2-131 b | 1 | 6.5 | 1.6 | -1.6 | 1.81 | 0.16 | -0.12 | 0 | Dai_2017 | Dai_2017 | Mayo_2018 |
| 11 | 0.2803 | 0 | 4599 | 0.71 | 0.68 | K2-141 b | 1 | 5.08 | 0.41 | -0.41 | 1.51 | 0.05 | -0.05 | -0.06 | Malavolta_2018 | Malavolta_2018 | Barragan_2018 |
| 12 | 0.8375 | 0 | 5599 | 0.91 | 1.09 | K-10 b | 1 | 4.61 | 1.27 | -1.46 | 1.489 | 0.07 | -0.062 | -0.15 | Esteves_2015 | Berger_2018 | Dumusque_2014 |
| 13 | 3.6961 | 0 | 5508 | 0.95 | 0.89 | K-20 b | 1 | 9.7 | 1.41 | -1.44 | 1.707 | 0.08 | -0.068 | 0.07 | Buchhave_2016 | Berger_2018 | Gautier_2012 |
| 14 | 2.78578 | 0 | 6202 | 1.41 | 1.93 | K-21 b | 1 | 5.08 | 1.72 | -1.72 | 1.639 | 0.019 | -0.015 | -0.1 | Lopez-Morales_2016 | Lopez-Morales_2016 | Howell_2012 |
| 15 | 2.42629 | 0 | 5656 | 1.07 | 1.07 | K-406 b | 1 | 6.35 | 1.4 | -1.4 | 1.558 | 0.148 | -0.119 | 0.26 | Marcy_2014 | Berger_2018 | Morton_2016 |
| 16 | 4.72674 | 0 | 5594 | 0.91 | 0.94 | K-93 b | 1 | 4.02 | 0.68 | -0.68 | 1.642 | 0.069 | -0.108 | -0.2 | Dressing_2015 | Berger_2018 | Ballard_2014 |
| 17 | 4.60358 | 0 | 4852 | 0.79 | 0.74 | K-99 b | 1 | 6.15 | 1.3 | -1.3 | 1.511 | 0.073 | -0.069 | 0.18 | Marcy_2014 | Berger_2018 | Morton_2016 |
| 18 | 24.7371 | 0 | 3216 | 0.179 | 0.2139 | LHS 1140 b | 1 | 6.98 | 0.98 | -0.98 | 1.727 | 0.032 | -0.032 | -0.24 | Ment_2018 | Ment_2018 | Dittmann_2017 |
| 19 | 0.78964 | 0 | 5576 | 1.11 | 1.16 | WASP-47 e | 1 | 6.83 | 0.66 | -0.66 | 1.81 | 0.027 | -0.027 | 0.36 | Vanderburg_2017 | Vanderburg_2017 | Almenara_2016 |
| 20 | 3.50473 | 1 | 5441 | 0.97 | 0.92 | K-18 b | 1 | 6.9 | 3.4 | -3.4 | 1.765 | 0.185 | -0.165 | 0.19 | Cochran_2011 | Berger_2018 | Morton_2016 |

**RV planets 2.0-3.0 Earth radii**

| | | | | | | | | | | | | | | | | |
|---|---|---|---|---|---|---|---|---|---|---|---|---|---|---|---|---|
| 1 | 41.6855 | 0 | 5766 | 1.67 | 1.08 | K2-56 b | 1 | 16.3 | 6 | -6.1 | 2.6 | 0.1 | -0.1 | -0.15 | Espinoza_2016 | Stassun_2017 | Mayo_2018 |
| 2 | 1.5804 | 0 | 3026 | 0.15 | 0.22 | GJ 1214 b | 1 | 6.26 | 0.86 | -0.86 | 2.85 | 0.2 | -0.2 | 0.15 | Harpsoe_2013 | Harpsoe_2013 | Charbonneau_2009 |
| 3 | 9.4909 | 0 | 5175 | 0.75 | 0.74 | HD 97658 b | 1 | 7.86 | 0.73 | -0.73 | 2.4 | 0.1 | -0.1 | -0.3 | Dragomir_2013 | Stassun_2017 | Van_Grootel_2014 |
| 4 | 9.1205 | 0 | 5089 | 0.78 | 0.72 | HIP 116454 b | 1 | 11.82 | 1.33 | -1.33 | 2.5 | 0.1 | -0.1 | -0.16 | Vanderburg_2015 | Stassun_2017 | |
| 5 | 10.0545 | 0 | 3896 | 0.6 | 0.56 | K2-3 b | 1 | 6.6 | 1.1 | -1.1 | 2.29 | 0.23 | -0.23 | -0.01 | Damasso_2018 | Damasso_2018 | Sinukoff_2016 |
| 6 | 13.8638 | 0 | 5010 | 0.74 | 0.71 | K2-110 b | 1 | 16.7 | 3.2 | -3.2 | 2.6 | 0.1 | -0.1 | -0.34 | Osborn_2017 | Osborn_2017 | Mayo_2018 |
| 7 | 32.9396 | 0 | 3457 | 0.36 | 0.41 | K2-18 b | 1 | 7.96 | 1.91 | -1.91 | 2.38 | 0.22 | -0.22 | 0.123 | Cloutier_2017 | Cloutier_2017 | Benneke_2017 |
| 8 | 45.2943 | 1 | 5599 | 0.91 | 1.09 | K-10 c | 1 | 17.2 | 1.9 | -1.9 | 2.35 | 0.1 | -0.1 | -0.15 | Dumusque_2014 | Berger_2018 | Weiss_2016 |
| 9 | 16.1457 | 0 | 4909 | 0.81 | 0.73 | K-102 e | 1 | 8.93 | 2 | -2 | 2.43 | 0.1 | -0.2 | 0.18 | Marcy_2014 | Berger_2018 | Morton_2016 |
| 10 | 13.5708 | 0 | 5860 | 1 | 1.04 | K-106 c | 1 | 10.44 | 3.2 | -3.2 | 2.5 | 0.32 | -0.32 | -0.12 | Marcy_2014 | Marcy_2014 | Morton_2016 |
| 11 | 43.8444 | 0 | 5860 | 1 | 1.04 | K-106 e | 1 | 11.17 | 5.8 | -5.8 | 2.56 | 0.33 | -0.33 | -0.12 | Marcy_2014 | Marcy_2014 | Morton_2016 |
| 12 | 16.092 | 0 | 5787 | 1.02 | 1 | K-131 b | 1 | 16.13 | 3.5 | -3.5 | 2.4 | 0.2 | -0.2 | 0.12 | Marcy_2014 | Marcy_2014 | Morton_2016 |
| 13 | 9.287 | 0 | 5520 | 0.936 | 0.88 | K-19 b | 1 | 8.4 | 1.6 | -1.5 | 2.3 | 0.1 | -0.1 | -0.08 | Malavolta_2017 | Berger_2018 | Ballard_2011 |
| 14 | 77.6113 | 0 | 5508 | 0.95 | 0.89 | K-20 d | 1 | 10.07 | 3.97 | -3.7 | 2.5 | 0.1 | -0.1 | 0.07 | Buchhave_2016 | Berger_2018 | Gautier_2012 |
| 15 | 6.2385 | 1 | 6285 | 1.19 | 1.36 | K-25 b | 1 | 9.6 | 4.2 | -4.2 | 2.73 | 0.12 | -0.11 | -0.04 | Marcy_2014 | Berger_2018 | Hadden_2014 |
| 16 | 10.5738 | 0 | 5678 | 1.03 | 1.08 | K-454 b | 1 | 6.84 | 1.4 | -1.4 | 2.06 | 0.1 | -0.2 | 0.27 | Gettel_2016 | Berger_2018 | Stassun_2017 |
| 17 | 9.67395 | 1 | 5235 | 0.88 | 0.85 | K-48 c | 1 | 14.61 | 2.3 | -2.3 | 2.56 | 0.1 | -0.1 | 0.17 | Marcy_2014 | Berger_2018 | Hadden_2014 |
| 18 | 5.39875 | 0 | 5789 | 1.08 | 1.26 | K-68 b | 1 | 8.3 | 2.2 | -2.4 | 2.32 | 0.1 | -0.1 | 0.12 | Gilliland_2013 | Berger_2018 | van_Eylen_2015 |
| 19 | 16.2385 | 0 | 5751 | 1 | 0.95 | K-96 b | 1 | 8.46 | 3.4 | -3.4 | 2.64 | 0.1 | -0.24 | 0.04 | Marcy_2014 | Berger_2018 | Morton_2016 |
| 20 | 4.754 | 0 | 4791 | 0.75 | 0.75 | K-113 b | 1 | 11.7 | 4.2 | -4.2 | 2.16 | 0.1 | -0.2 | 0.05 | Marcy_2014 | Berger_2018 | Morton_2016 |
| 21 | 10.8541 | 0 | 5508 | 0.95 | 0.89 | K-20 c | 1 | 12.75 | 2.17 | -2.24 | 2.87 | 0.19 | -0.13 | 0.07 | Buchhave_2016 | Berger_2018 | Gautier_2012 |
| 22 | 3.47175 | 0 | 4975 | 0.83 | 0.787 | EPIC246471491 b | 1 | 9.68 | 1.21 | -1.37 | 2.59 | 0.06 | -0.06 | 0.0 | Palle_2018 | Palle_2018 | |
| 23 | 50.81895 | 0 | 5372 | 0.88 | 0.85 | K2-263 b | 1 | 14.8 | 3.1 | -3.1 | 2.41 | 0.12 | -0.12 | -0.08 | Mortier_2018 | Mortier_2018 | |
| 24 | 6.2682 | 0 | 6037 | 1.094 | 1.10 | π Mensae c | 1 | 4.82 | 0.84 | -0.86 | 2.14 | 0.044 | -0.044 | 0.08 | Huang_2018 | Huang_2018 | Gandolfi_2018 |
| **RV planets 3.0-4.0 Earth radii** | | | | | | | | | | | | | | | | | |
| 1 | 29.8454 | 0 | 5261 | 0.86 | 0.86 | HD 3167 c | 1 | 9.8 | 1.3 | -1.24 | 3.01 | 0.42 | -0.28 | 0.04 | Christiansen_2017 | Christiansen_2017 | |
| 2 | 10.9542 | 1 | 5466 | 0.956 | 0.906 | K-88 b | 1 | 8.7 | 2.5 | -2.5 | 3.82 | 0.17 | -0.34 | 0.2 | Nesvorny_2013 | Berger_2018 | |
| 3 | 2.50806 | 0 | 4728 | 0.81 | 0.747 | K-94 b | 1 | 10.84 | 1.4 | -1.4 | 3.19 | 0.14 | -0.25 | 0.34 | Marcy_2018 | Berger_2018 | |
| 4 | 11.5231 | 0 | 5654 | 1.08 | 1.418 | K-95 b | 1 | 13 | 2.9 | -2.9 | 3.14 | 0.14 | -0.13 | 0.3 | Marcy_2018 | Berger_2018 | |
| 5 | 7.138048 | 0 | 4975 | 0.83 | 0.787 | EPIC246471491 c | 1 | 15.68 | 2.28 | -2.13 | 3.53 | 0.08 | -0.08 | 0.0 | Palle_2018 | Palle_2018 | |
| 6 | 1.33735 | 0 | 5143 | 0.75 | 0.84 | NGTS-4 b | 1 | 20.6 | 3.0 | -3.0 | 3.18 | 0.26 | -0.26 | -0.28 | West_2018 | West_2018 | |
| **RV planets 4.0-8.0 Earth radii** | | | | | | | | | | | | | | | | | |
| 1 | 3.33665 | 0 | 3600 | 0.51 | 0.547 | GJ 3470 b | 1 | 13.9 | 1.5 | -1.5 | 4.57 | 0.18 | -0.18 | 0.18 | Awlphan_2016 | Awlphan_2016 | |
| 2 | 2.8492725 | 0 | 4300 | 0.688 | 0.715 | K2-55 b | 1 | 43.13 | 5.98 | -5.8 | 4.41 | 0.32 | -0.28 | 0.376 | Dressing_2018 | Dressing_2018 | |
| 3 | 3.185315 | 0 | 4985 | 0.85 | 0.81 | HATS-7 b | 1 | 38.1396 | 3.81396 | -3.81396 | 6.311 | 0.516 | -0.381 | 0.25 | Bakos_2015 | Bakos_2015 | |
| 4 | 11.81399 | 0 | 5499 | 1.12 | 1.66 | HD 89345 b | 1 | 35.7 | 3.3 | -3.3 | 6.86 | 0.14 | -0.14 | 0.45 | Van_Eylen_2018 | Van_Eylen_2018 | |
| 5 | 4.73401 | 0 | 5474 | 1.12 | 1.75 | K2-108 b | 1 | 59.4 | 4.4 | -4.4 | 5.28 | 0.54 | -0.54 | 0.33 | Petigura_2017 | Petigura_2017 | Mayo_2018 |
| 6 | 8.99213 | 0 | 5275 | 0.86 | 0.84 | K2-32 b | 1 | 16.5 | 2.7 | -2.7 | 5.13 | 0.28 | -0.28 | -0.02 | Petigura_2017 | Petigura_2017 | Dai_2016 |
| 7 | 10.13675 | 0 | 6120 | 1.07 | 1.31 | K2-98 b | 1 | 32.2 | 8.1 | -8.1 | 4.3 | 0.3 | -0.2 | -0.2 | Barragan_2016 | Barragan_2016 | Mayo_2018 |
| 8 | 3.836169 | 0 | 4910 | 0.84 | 0.76 | WASP-156 b | 1 | 40.68224 | 3.1783 | -2.86047 | 5.717 | 0.224 | -0.224 | 0.24 | Demangeon_2017 | Demangeon_2017 | |
| 9 | 4.88782 | 0 | 4708 | 0.88 | 0.77 | HAT-P-11 b | 1 | 23.4 | 1.5 | -1.5 | 4.36 | 0.06 | -0.06 | 0.31 | Yee_2018 | Yee_2018 | Bakos_2010 |
| 10 | 3.21346 | 0 | 5857 | 1.22 | 1.49 | K-4 b | 1 | 24.472 | 3.814 | -3.814 | 4.076 | 0.181 | -0.17 | 0.17 | Borucki_2010 | Berger_2018 | Southworth_2011 |
| 11 | 6.21229 | 0 | 5080 | 0.88 | 0.77 | CoRoT-8 b | 1 | 69.92 | 9.53 | -9.53 | 6.39 | 0.22 | -0.22 | 0.3 | Borde_2010 | Borde_2010 | Southworth_2011 |
| 12 | 4.23452 | 0 | 5079 | 1.12 | 0.87 | HAT-P-26 b | 1 | 22.2481 | 6.3566 | -6.3566 | 7.062 | 0.448 | -0.448 | -0.04 | Stassun_2017 | Stassun_2017 | Stevenson_2016 |
| 13 | 7.64159 | 1 | 5441 | 0.97 | 0.916 | K-18 c | 1 | 17.3 | 1.9 | -1.9 | 4.263 | 0.183 | -0.172 | 0.19 | Cochran_2011 | Berger_2018 | Hadden_2014 |
| 14 | 14.85888 | 1 | 5441 | 0.97 | 0.916 | K-18 d | 1 | 16.4 | 1.4 | -1.4 | 5.16 | 0.219 | -0.215 | 0.19 | Cochran_2011 | Berger_2018 | Hadden_Lithwich_2017 |
| 15 | 12.7204 | 1 | 6285 | 1.19 | 1.36 | K-25 c | 1 | 24.6 | 5.7 | -5.7 | 5.134 | 0.216 | -0.414 | -0.04 | Marcy_2014 | Berger_2018 | Hadden_2014 |
| 16 | 2.64388312 | 0 | 3416 | 0.47 | 0.455 | GJ 436 b | 1 | 21.36 | 0.2 | -0.2 | 4.191 | 0.109 | -0.109 | 0.02 | Trifonov_2018 | Turner_2016 | Dressing_2018 Maciejewski_2014 |
| **TTV planets <8.0 Earth radii** | | | | | | | | | | | | | | | | | |
| 1 | 10.3039 | 1 | 5763 | 1.04 | 1.091 | K-11 b | 0 | 2.78 | 0.64 | -0.66 | 1.916 | 0.095 | -0.081 | -0.041 | Bedell_2017 | Berger_2018 | |
| 2 | 13.0241 | 1 | 5763 | 1.04 | 1.091 | K-11 c | 0 | 5 | 1.3 | -1.35 | 3.006 | 0.138 | -0.124 | -0.041 | Bedell_2017 | Berger_2018 | |
| 3 | 22.6845 | 1 | 5763 | 1.04 | 1.091 | K-11 d | 0 | 8.13 | 0.67 | -0.66 | 3.338 | 0.158 | -0.138 | -0.041 | Bedell_2017 | Berger_2018 | |
| 4 | 31.9996 | 1 | 5763 | 1.04 | 1.091 | K-11 e | 0 | 9.48 | 0.86 | -0.88 | 4.3 | 0.188 | -0.309 | -0.041 | Bedell_2017 | Berger_2018 | |
| 5 | 46.6888 | 1 | 5836 | 1.04 | 1.091 | K-11 f | 0 | 2.53 | 0.49 | -0.45 | 2.627 | 0.124 | -0.116 | -0.041 | Bedell_2017 | Berger_2018 | |
| 6 | 12.28 | 1 | 3884 | 0.54 | 0.607 | K-26 b | 0 | 5.12 | 0.65 | -0.61 | 3.197 | 0.097 | -0.093 | -0.13 | Jontof_Hutter_2016 | Berger_2018 | |
| 7 | 17.2559 | 1 | 3884 | 0.54 | 0.607 | K-26 c | 0 | 6.2 | 0.65 | -0.65 | 2.982 | 0.278 | -0.229 | -0.13 | Jontof_Hutter_2016 | Berger_2018 | |
| 8 | 10.3384 | 1 | 5378 | 0.98 | 0.749 | K-29 b | 0 | 4.51 | 1.41 | -1.47 | 2.564 | 0.16 | -0.141 | -0.04 | Jontof_Hutter_2016 | Berger_2018 | |
| 9 | 13.2884 | 1 | 5378 | 0.98 | 0.749 | K-29 c | 0 | 4 | 1.23 | -1.29 | 2.396 | 0.151 | -0.133 | -0.04 | Jontof_Hutter_2016 | Berger_2018 | |
| 10 | 29.33434 | 1 | 5454 | 0.99 | 0.834 | K-30 b | 0 | 8.8 | 0.6 | -0.5 | 1.837 | 0.124 | -0.171 | 0.18 | Hadden_Lithwick_2017 | Berger_2018 | |
| 11 | 13.83989 | 1 | 5979 | 1.07 | 1.652 | K-36 b | 0 | 3.9 | 0.2 | -0.22 | 1.514 | 0.082 | -0.071 | -0.2 | Hadden_Lithwick_2017 | Berger_2018 | |
| 12 | 16.23855 | 1 | 5979 | 1.07 | 1.652 | K-36 c | 0 | 7.5 | 0.3 | -0.3 | 3.688 | 0.165 | -0.153 | -0.2 | Hadden_Lithwick_2017 | Berger_2018 | |
| 13 | 45.154 | 1 | 5674 | 1.04 | 0.833 | K-51 b | 0 | 2.3 | 1.7 | -1.6 | 6.559 | 0.316 | -0.297 | null | Hadden_Lithwick_2017 | Berger_2018 | |
| 14 | 7.1334 | 1 | 5834 | 1.041 | 1.469 | K-60 b | 0 | 3.7 | 0.6 | -0.6 | 1.845 | 0.339 | -0.117 | -0.09 | Hadden_Lithwick_2017 | Berger_2018 | |
| 15 | 8.9187 | 1 | 5834 | 1.041 | 1.469 | K-60 c | 0 | 2 | 0.3 | -0.5 | 2.062 | 0.378 | -0.131 | -0.09 | Hadden_Lithwick_2017 | Berger_2018 | |
| 16 | 11.8981 | 1 | 5834 | 1.041 | 1.469 | K-60 d | 0 | 3.9 | 0.7 | -0.6 | 1.901 | 0.495 | -0.17 | -0.09 | Hadden_Lithwick_2017 | Berger_2018 | |
| 17 | 27.4029 | 1 | 6389 | 1.17 | 1.345 | K-79 c | 0 | 5.9 | 1.9 | -2.3 | 3.684 | 0.181 | -0.165 | -0.073 | Jontof_Hutter_2014 | Berger_2018 | |
| 18 | 52.0902 | 1 | 6389 | 1.17 | 1.345 | K-79 d | 0 | 6 | 2.1 | -1.6 | 7.251 | 0.34 | -0.322 | -0.073 | Jontof_Hutter_2014 | Berger_2018 | |
| 19 | 81.0659 | 1 | 6389 | 1.17 | 1.345 | K-79 e | 0 | 4.1 | 1.2 | -1.1 | 3.054 | 0.76 | -0.186 | -0.073 | Jontof_Hutter_2014 | Berger_2018 | |
| 20 | 7.05246 | 1 | 4665 | 0.73 | 0.681 | K-80 b | 0 | 6.93 | 1.05 | -0.7 | 2.592 | 0.116 | -0.11 | 0.04 | MacDonald_2016 | Berger_2018 | |
| 21 | 9.52355 | 1 | 4665 | 0.73 | 0.681 | K-80 c | 0 | 6.74 | 1.23 | -0.86 | 2.594 | 0.121 | -0.109 | 0.04 | MacDonald_2016 | Berger_2018 | |
| 22 | 3.07222 | 1 | 4665 | 0.73 | 0.681 | K-80 d | 0 | 6.75 | 0.69 | -0.51 | 1.561 | 0.081 | -0.064 | 0.04 | MacDonald_2016 | Berger_2018 | |
| 23 | 4.64489 | 1 | 4665 | 0.73 | 0.681 | K-80 e | 0 | 4.13 | 0.81 | -0.95 | 1.602 | 0.081 | -0.069 | 0.04 | MacDonald_2016 | Berger_2018 | |
| 24 | 10.4208 | 1 | 5559 | 0.91 | 0.936 | K-307 b | 0 | 8.8 | 0.9 | -0.9 | 2.85 | 0.274 | -0.208 | 0.19 | Hadden_Lithwick_2017 | Berger_2018 | |
| 25 | 13.0729 | 1 | 5559 | 0.91 | 0.936 | K-307 c | 0 | 3.9 | 0.7 | -0.7 | 2.569 | 0.279 | -0.211 | 0.19 | Hadden_Lithwick_2017 | Berger_2018 | |
| 26 | 66.0634 | 1 | 5989 | 1.08 | 1.013 | K-289 d | 0 | 4 | 0.9 | -0.9 | 2.668 | 0.17 | -0.17 | 0.05 | Rowe_2014 | Rowe_2014 | |
| 27 | 1.51087081 | 1 | 2559 | 0.08 | 0.12 | TRAPPIST-1 b | 0 | 1.22 | 0.15 | -0.15 | 1.121 | 0.031 | -0.032 | 0.04 | Demory_2018 | Grimm_2018 | Delrez_2018 |
| 28 | 2.4218233 | 1 | 2559 | 0.08 | 0.12 | TRAPPIST-1 c | 0 | 1.24 | 0.15 | -0.15 | 1.095 | 0.03 | -0.031 | 0.04 | Demory_2018 | Grimm_2018 | Delrez_2018 |
| 29 | 4.04961 | 1 | 2559 | 0.08 | 0.12 | TRAPPIST-1 d | 0 | 0.37 | 0.04 | -0.04 | 0.784 | 0.023 | -0.023 | 0.04 | Demory_2018 | Grimm_2018 | Delrez_2018 |
| 30 | 6.099615 | 1 | 2559 | 0.08 | 0.12 | TRAPPIST-1 e | 0 | 0.66 | 0.079 | -0.075 | 0.91 | 0.026 | -0.027 | 0.04 | Demory_2018 | Grimm_2018 | Delrez_2018 |
| 31 | 9.20669 | 1 | 2559 | 0.08 | 0.12 | TRAPPIST-1 f | 0 | 0.97 | 0.08 | -0.078 | 1.046 | 0.029 | -0.03 | 0.04 | Demory_2018 | Grimm_2018 | Delrez_2018 |
| 32 | 12.35294 | 1 | 2559 | 0.08 | 0.12 | TRAPPIST-1 g | 0 | 1.27 | 0.098 | -0.095 | 1.148 | 0.032 | -0.033 | 0.04 | Demory_2018 | Grimm_2018 | Delrez_2018 |
| 33 | 18.767 | 1 | 2559 | 0.08 | 0.12 | TRAPPIST-1 h | 0 | 0.37 | 0.049 | -0.056 | 0.773 | 0.027 | -0.026 | 0.04 | Demory_2018 | Grimm_2018 | Delrez_2018 |

## References for SI reference citations

1. R. L. Akeson *et al.*, The NASA Exoplanet Archive: Data and Tools for Exoplanet Research. *Publ. Astron. Soc. Pacific*. **125**, 989 (2013).
2. S. E. Thompson *et al.*, Planetary Candidates Observed by Kepler. VIII. A Fully Automated Catalog With Measured Completeness and Reliability Based on Data Release 25. *Astrophys. J. Suppl. Ser. Vol. 235, Issue 2, Artic. id. 38, 49 pp. (2018).* **235** (2017), doi:10.3847/1538-4365/aab4f9.
3. K. G. Stassun, E. Corsaro, J. A. Pepper, B. S. Gaudi, Empirical Accurate Masses and Radii of Single Stars with *TESS* and *Gaia*. *Astron. J.* **155**, 22 (2017).
4. T. A. Berger, D. Huber, E. Gaidos, J. L. van Saders, Revised Radii of Kepler Stars and Planets using Gaia Data Release 2 (2018) (available at http://arxiv.org/abs/1805.00231).
5. C. M. Persson *et al.*, An 8 Mearth super-Earth in a 2.2 day orbit around the K5V star K2-216 (2018) (available at http://arxiv.org/abs/1805.04774).
6. A. W. Mayo *et al.*, 275 Candidates and 149 Validated Planets Orbiting Bright Stars in K2 Campaigns 0-10. *Astron. Journal, Vol. 155, Issue 3, Artic. id. 136, 25 pp. (2018).* **155** (2018), doi:10.3847/1538-3881/aaadff.
7. B. O. Demory *et al.*, A map of the large day-night temperature gradient of a super-Earth exoplanet. *Nature*. **532** (2016), doi:10.1038/nature17169.
8. J. ~N. Winn *et al.*, A Super-Earth Transiting a Naked-eye Star. *\apjl*. **737**, L18 (2011).
9. S. ~C. ~C. Barros *et al.*, Revisiting the transits of CoRoT-7b at a lower activity level. *\aap*. **569**, A74 (2014).
10. R. ~D. Haywood *et al.*, Planets and stellar activity: hide and seek in the CoRoT-7 system. *\mnras*. **443**, 2517–2531 (2014).
11. M. Gillon *et al.*, Two massive rocky planets transiting a K-dwarf 6.5 parsecs away (2017), doi:10.1038/s41550-017-0056.
12. F. Motalebi *et al.*, The HARPS-N Rocky Planet Search I. HD219134b: A transiting rocky planet in a multi-planet system at 6.5 pc from the Sun. *Astron. Astrophys. Vol. 584, id.A72, 12 pp.* **584** (2015), doi:10.1051/0004-6361/201526822.
13. J. L. Christiansen *et al.*, Three's Company: An Additional Non-transiting Super-Earth in the Bright HD 3167 System, and Masses for All Three Planets. *Astron. J.* **154**, 122 (2017).
14. D. Gandolfi *et al.*, The Transiting Multi-planet System HD 3167: A 5.7 $M_{\oplus}$ Super-Earth and an 8.3 $M_{\oplus}$ Mini-Neptune. *Astron. J.* **154**, 123 (2017).
15. M. Damasso *et al.*, Eyes on K2-3: A system of three likely sub-Neptunes characterized with HARPS-N and HARPS (2018) (available at http://arxiv.org/abs/1802.08320).
16. E. Sinukoff *et al.*, Eleven Multi-planet Systems from K2 Campaigns 1 & 2 and the Masses of Two Hot Super-Earths. *Astrophys. Journal, Vol. 827, Issue 1, Artic. id. 78, 27 pp. (2016).* **827** (2015), doi:10.3847/0004-637X/827/1/78.
17. E. W. Guenther *et al.*, K2-106, a system containing a metal-rich planet and a planet of lower density (2017), doi:10.1051/0004-6361/201730885.
18. E. R. Adams *et al.*, Ultra Short Period Planets in K2 with companions: a double transiting system for EPIC 220674823. *Astron. Journal, Vol. 153, Issue 2, Artic. id. 82, 7 pp. (2017).* **153** (2016), doi:10.3847/1538-3881/153/2/82.
19. E. Sinukoff *et al.*, K2-66b and K2-106b: Two extremely hot sub-Neptune-size planets with high densities. *Astron. Journal, Vol. 153, Issue 6, Artic. id. 271, 13 pp. (2017).* **153** (2017), doi:10.3847/1538-3881/aa725f.
20. F. Dai *et al.*, The Discovery and Mass Measurement of a New Ultra-short-period Planet:

K2-131b. *Astron. J.* **154**, 226 (2017).
21. L. Malavolta *et al.*, An ultra-short period rocky super-Earth with a secondary eclipse and a Neptune-like companion around K2-141 (2018), doi:10.3847/1538-3881/aaa5b5.
22. O. Barragán *et al.*, K2-141 b: A 5-M$_\oplus$ super-Earth transiting a K7 V star every 6.7 hours. *Astron. Astrophys. Vol. 612, id.A95, 11 pp.* **612** (2017), doi:10.1051/0004-6361/201732217.
23. L. ~J. Esteves, E. ~J. ~W. De Mooij, R. Jayawardhana, Changing Phases of Alien Worlds: Probing Atmospheres of Kepler Planets with High-precision Photometry. *\apj*. **804**, 150 (2015).
24. X. Dumusque *et al.*, The Kepler-10 Planetary System Revisited by HARPS-N: A Hot Rocky World and a Solid Neptune-Mass Planet. *\apj*. **789**, 154 (2014).
25. G. W. Marcy *et al.*, Masses, Radii, and Orbits of Small Kepler Planets: The Transition from Gaseous to Rocky Planets. **20** (2014), doi:10.1088/0067-0049/210/2/20.
26. L. A. Buchhave *et al.*, A 1.9 Earth radius rocky planet and the discovery of a non-transiting planet in the Kepler-20 system. *Astron. Journal, Vol. 152, Issue 6, Artic. id. 160, 12 pp. (2016).* **152** (2016), doi:10.3847/0004-6256/152/6/160.
27. T. N. Gautier *et al.*, Kepler-20: A Sun-like Star with Three Sub-Neptune Exoplanets and Two Earth-size Candidates. *Astrophys. J.* **749**, 15 (2012).
28. M. Lopez-Morales *et al.*, Kepler-21b: A rocky planet around a V = 8.25 magnitude star. *ArXiv e-prints* (2016).
29. S. B. Howell *et al.*, Kepler-21b: A 1.6 R Earth Planet Transiting the Bright Oscillating F Subgiant Star HD 179070. *Astrophys. J.* **746**, 123 (2012).
30. T. D. Morton *et al.*, False positive probabilties for all Kepler Objects of Interest: 1284 newly validated planets and 428 likely false positives. *Astrophys. Journal, Vol. 822, Issue 2, Artic. id. 86, 15 pp. (2016).* **822** (2016), doi:10.3847/0004-637X/822/2/86.
31. C. ~D. Dressing *et al.*, The Mass of Kepler-93b and The Composition of Terrestrial Planets. *\apj*. **800**, 135 (2015).
32. S. Ballard *et al.*, Kepler-93b: A Terrestrial World Measured to within 120 km, and a Test Case for a New Spitzer Observing Mode. *Astrophys. J.* **790**, 12 (2014).
33. W. J. Borucki *et al.*, Kepler-22b: A 2.4 Earth-radius Planet in the Habitable Zone of a Sun-like Star. *Astrophys. J.* **745**, 120 (2012).
34. J. A. Dittmann *et al.*, A temperate rocky super-Earth transiting a nearby cool star. *Nature, Vol. 544, Issue 7650, pp. 333-336 (2017).* **544**, 333–336 (2017).
35. A. Vanderburg *et al.*, Precise Masses in the WASP-47 System (2017), doi:10.3847/1538-3881/aa918b.
36. J. M. Almenara, R. F. Díaz, X. Bonfils, S. Udry, Absolute densities, masses, and radii of the WASP-47 system determined dynamically. *Astron. Astrophys. Vol. 595, id.L5, 12 pp.* **595** (2016), doi:10.1051/0004-6361/201629770.
37. X. Bonfils *et al.*, Radial velocity follow-up of GJ1132 with HARPS. A precise mass for planet "b" and the discovery of a second planet (2018) (available at http://arxiv.org/abs/1806.03870).
38. N. Espinoza *et al.*, Discovery and Validation of a High-Density sub-Neptune from the K2 Mission. *Astrophys. Journal, Vol. 830, Issue 1, Artic. id. 43, 11 pp. (2016).* **830** (2016), doi:10.3847/0004-637X/830/1/43.
39. K. B. W. Harpsøe *et al.*, The Transiting System GJ1214: High-Precision Defocused Transit Observations and a Search for Evidence of Transit Timing Variation. *Astron.*

*Astrophys. Vol. 549, id.A10, 10 pp.* **549** (2012), doi:10.1051/0004-6361/201219996.

40. D. Charbonneau *et al.*, A super-Earth transiting a nearby low-mass star. *\nat*. **462**, 891–894 (2009).
41. D. Dragomir *et al.*, MOST Detects Transits of HD 97658b, a Warm, Likely Volatile-rich Super-Earth. *\apjl*. **772**, L2 (2013).
42. V. Van Grootel *et al.*, Transit confirmation and improved stellar and planet parameters for the super-Earth HD 97658 b and its host star. *Astrophys. Journal, Vol. 786, Issue 1, Artic. id. 2, 11 pp. (2014).* **786** (2014), doi:10.1088/0004-637X/786/1/2.
43. A. Vanderburg *et al.*, Characterizing K2 Planet Discoveries: A Super-Earth Transiting the Bright K Dwarf HIP{\nbsp}116454. *\apj*. **800**, 59 (2015).
44. H. P. Osborn *et al.*, K2-110 b - a massive mini-Neptune exoplanet. *Astron. Astrophys. Vol. 604, id.A19, 8 pp.* **604** (2016), doi:10.1051/0004-6361/201628932.
45. R. Cloutier *et al.*, Characterization of the K2-18 multi-planetary system with HARPS: A habitable zone super-Earth and discovery of a second, warm super-Earth on a non-coplanar orbit. *Astron. Astrophys. Vol. 608, id.A35, 13 pp.* **608** (2017), doi:10.1051/0004-6361/201731558.
46. B. Benneke *et al.*, Spitzer Observations Confirm and Rescue the Habitable-Zone Super-Earth K2-18b for Future Characterization. *Astrophys. Journal, Vol. 834, Issue 2, Artic. id. 187, 10 pp. (2017).* **834** (2016), doi:10.3847/1538-4357/834/2/187.
47. L. M. Weiss *et al.*, Revised Masses and Densities of the Planets around Kepler-10. *Astrophys. Journal, Vol. 819, Issue 1, Artic. id. 83, 22 pp. (2016).* **819** (2016), doi:10.3847/0004-637X/819/1/83.
48. L. Malavolta *et al.*, The Kepler-19 System: A Thick-envelope Super-Earth with Two Neptune-mass Companions Characterized Using Radial Velocities and Transit Timing Variations. *Astron. Journal, Vol. 153, Issue 5, Artic. id. 224, 14 pp. (2017).* **153** (2017), doi:10.3847/1538-3881/aa6897.
49. S. Ballard *et al.*, The Kepler-19 System: A Transiting 2.2 R ⊕ Planet and a Second Planet Detected via Transit Timing Variations. *Astrophys. Journal, Vol. 743, Issue 2, Artic. id. 200, 20 pp. (2011).* **743** (2011), doi:10.1088/0004-637x/743/2/200.
50. S. Hadden, Y. Lithwick, Densities and Eccentricities of 139 Kepler Planets from Transit Time Variations. *Astrophys. Journal, Vol. 787, Issue 1, Artic. id. 80, 7 pp. (2014).* **787** (2014), doi:10.1088/0004-637x/787/1/80.
51. S. Gettel *et al.*, The Kepler-454 System: A Small, Not-rocky Inner Planet, a Jovian World, and a Distant Companion. *Astrophys. Journal, Vol. 816, Issue 2, Artic. id. 95, 13 pp. (2016).* **816** (2015), doi:10.3847/0004-637X/816/2/95.
52. R. L. Gilliland *et al.*, Kepler-68: Three Planets, One with a Density between that of Earth and Ice Giants. *Astrophys. J.* **766**, 40 (2013).
53. V. Van Eylen, S. Albrecht, Eccentricity from transit photometry: small planets in Kepler multi-planet systems have low eccentricities. *Astrophys. Journal, Vol. 808, Issue 2, Artic. id. 126, 20 pp. (2015).* **808** (2015), doi:10.1088/0004-637X/808/2/126.
54. G. Á. Bakos *et al.*, HATS-7b: A Hot Super Neptune Transiting a Quiet K Dwarf Star. *Astrophys. Journal, Vol. 813, Issue 2, Artic. id. 111, 10 pp. (2015).* **813** (2015), doi:10.1088/0004-637X/813/2/111.
55. S. W. Yee *et al.*, HAT-P-11: Discovery of a Second Planet and a Clue to Understanding Exoplanet Obliquities. *Astron. Journal, Vol. 155, Issue 6, Artic. id. 255, 13 pp. (2018).* **155** (2018), doi:10.3847/1538-3881/aabfec.

56. V. Van Eylen *et al.*, HD 89345: a bright oscillating star hosting a transiting warm Saturn-sized planet observed by K2. *Mon. Not. R. Astron. Soc. Vol. 478, Issue 4, p.4866-4880.* **478**, 4866–4880 (2018).
57. E. A. Petigura *et al.*, Four Sub-Saturns with Dissimilar Densities: Windows into Planetary Cores and Envelopes. *Astron. Journal, Vol. 153, Issue 4, Artic. id. 142, 19 pp. (2017).* **153** (2017), doi:10.3847/1538-3881/aa5ea5.
58. F. Dai *et al.*, Doppler Monitoring of five K2 Transiting Planetary Systems. *Astrophys. Journal, Vol. 823, Issue 2, Artic. id. 115, 16 pp. (2016).* **823** (2016), doi:10.3847/0004-637X/823/2/115.
59. O. Barragán *et al.*, K2-98 b: A 32-M$_\oplus$ Neptune-sized planet in a 10-day orbit transiting an F8 star. *Astron. Journal, Vol. 152, Issue 6, Artic. id. 193, 9 pp. (2016).* **152** (2016), doi:10.3847/0004-6256/152/6/193.
60. O. D. S. Demangeon *et al.*, The discovery of WASP-151b, WASP-153b, WASP-156b: Insights on giant planet migration and the upper boundary of the Neptunian desert. *Astron. Astrophys. Vol. 610, id.A63, 19 pp.* **610** (2017), doi:10.1051/0004-6361/201731735.
61. P. Bordé *et al.*, Transiting exoplanets from the CoRoT space mission. XI. CoRoT-8b: a hot and dense sub-Saturn around a K1 dwarf. *Astron. Astrophys. Vol. 520, id.A66, 10 pp.* **520** (2010), doi:10.1051/0004-6361/201014775.
62. J. Southworth, John, Homogeneous studies of transiting extrasolar planets. IV. Thirty systems with space-based light curves. *Mon. Not. R. Astron. Soc. Vol. 417, Issue 3, pp. 2166-2196.* **417**, 2166–2196 (2011).
63. K. B. Stevenson *et al.*, A Search for Water in the Atmosphere of HAT-P-26b Using LDSS-3C. *Astrophys. Journal, Vol. 817, Issue 2, Artic. id. 141, 10 pp. (2016).* **817** (2015), doi:10.3847/0004-637X/817/2/141.
64. T. Trifonov *et al.*, The CARMENES search for exoplanets around M dwarfs. First visual-channel radial-velocity measurements and orbital parameter updates of seven M-dwarf planetary systems. *Astron. Astrophys. Vol. 609, id.A117, 24 pp.* **609** (2017), doi:10.1051/0004-6361/201731442.
65. J. D. Turner *et al.*, Ground-based near-UV observations of 15 transiting exoplanets: Constraints on their atmospheres and no evidence for asymmetrical transits. *Mon. Not. R. Astron. Soc. Vol. 459, Issue 1, p.789-819*. **459**, 789–819 (2016).
66. G. Maciejewski *et al.*, On the GJ 436 planetary system. *Acta Astron. vol 64, no 4, p. 323-335*. **64**, 323–335 (2015).
67. T. Holczer *et al.*, Transit Timing Observations from Kepler. IX. Catalog of the Full Long-cadence Data Set. *apjs*. **225**, 9 (2016).
68. D. Jontof-Hutter *et al.*, Secure Mass Measurements from Transit Timing: 10 Kepler Exoplanets between 3 and 8 M⊕ with Diverse Densities and Incident Fluxes. *Astrophys. J.* **820**, 39 (2016).
69. S. Hadden, Y. Lithwick, Kepler Planet Masses and Eccentricities from Transit Timing Variations. *Am. Astron. Soc. AAS Meet. #229, id.401.03*. **229** (2017) (available at http://adsabs.harvard.edu/abs/2017AAS...22940103H).
70. S. L. Grimm *et al.*, The nature of the TRAPPIST-1 exoplanets (2018), doi:10.1051/0004-6361/201732233.
71. J. J. Lissauer *et al.*, A closely packed system of low-mass, low-density planets transiting Kepler-11. *Nature*. **470**, 53–58 (2011).
72. M. Bedell *et al.*, Kepler-11 is a Solar Twin: Revising the Masses and Radii of Benchmark

Planets via Precise Stellar Characterization. *Astrophys. J.* **839**, 94 (2017).
73. J. A. Carter *et al.*, Kepler-36: A Pair of Planets with Neighboring Orbits and Dissimilar Densities. *Science (80-. ).* **337**, 556–559 (2012).
74. K. Ment *et al.*, A second planet with an Earth-like composition orbiting the nearby M dwarf LHS 1140 (2018) (available at http://arxiv.org/abs/1808.00485).
75. E. Palle *et al.*, Detection and Doppler monitoring of EPIC 246471491, a system of four transiting planets smaller than Neptune (2018) (available at https://arxiv.org/abs/1808.00575).
76. L. Delrez *et al.*, Early 2017 observations of TRAPPIST-1 with Spitzer. *Mon. Not. R. Astron. Soc.* **475**, 3577–3597 (2018).
77. A. Mortier *et al.*, K2-263 b: a 50 d period sub-Neptune with a mass measurement using HARPS-N. *Mon. Not. R. Astron. Soc. Vol. 481, Issue 2, p.1839-1847.* **481**, 1839–1847 (2018).
78. C. X. Huang *et al.*, TESS Discovery of a Transiting Super-Earth in the $\Pi$ Mensae System (2018) (available at http://arxiv.org/abs/1809.05967).
79. D. Gandolfi *et al.*, TESS's first planet: a super-Earth transiting the naked-eye star $\pi$ Mensae (2018) (available at http://arxiv.org/abs/1809.07573).
80. R. G. West *et al.*, NGTS-4b: A sub-Neptune Transiting in the Desert (2018) (available at http://arxiv.org/abs/1809.00678).
81. J. N. Winn, in *Handbook of Exoplanets*, H. J. Deeg, J. . Belmonte, Eds. (Springer, 2018; http://arxiv.org/abs/1801.08543).
82. C. W. Ormel, B. Liu, D. Schoonenberg, Formation of TRAPPIST-1 and other compact systems. *Astron. Astrophys.* **604**, A1 (2017).
83. M. Gillon *et al.*, Seven temperate terrestrial planets around the nearby ultracool dwarf star TRAPPIST-1. *Nature*. **542**, 456–460 (2017).
84. F. Fressin *et al.*, Two Earth-sized planets orbiting Kepler-20. *\nat*. **482**, 195–198 (2012).
85. T. N. Gautier *et al.*, Kepler-20: A Sun-like Star with Three Sub-Neptune Exoplanets and Two Earth-size Candidates. *Astrophys. Journal, Vol. 749, Issue 1, Artic. id. 15, 19 pp. (2012).* **749** (2011), doi:10.1088/0004-637X/749/1/15.
86. A. S. Bonomo *et al.*, A giant impact as the likely origin of different twins in the Kepler-107 exoplanet system. *Nat. Astron.*, 1 (2019).
87. J. ~F. Rowe *et al.*, Validation of Kepler's Multiple Planet Candidates. III. Light Curve Analysis and Announcement of Hundreds of New Multi-planet Systems. *\apj*. **784**, 45 (2014).
88. S. Hadden, Y. Lithwick, Numerical and analytical modeling of transit timing variations. *ApJ*. **828**, 44 (2016).
89. S. Wang, D.-H. Wu, T. Barclay, G. P. Laughlin, Updated Masses for the TRAPPIST-1 Planets (2017) (available at http://arxiv.org/abs/1704.04290).
90. C. T. Unterborn, S. J. Desch, N. R. Hinkel, A. Lorenzo, Inward migration of the TRAPPIST-1 planets as inferred from their water-rich compositions. *Nat. Astron.* **2**, 297–302 (2018).
91. T. D. Morton, J. N. Winn, Obliquities of kepler stars: comparison of single- and multiple-transit systems. *Astrophys. J.* **796**, 47 (2014).
92. J. M. Brewer, S. Wang, D. A. Fischer, D. Foreman-Mackey, Compact Multi-planet Systems are more Common around Metal-poor Hosts. *Astrophys. J.* **867**, L3 (2018).
93. R. I. Dawson, E. J. Lee, E. Chiang, Correlations Between Compositions and Orbits

Established By the Giant Impact Era of Planet Formation. *Astrophys. J.* **822**, 54 (2016).
94. J. Moriarty, S. Ballard, The kepler dichotomy in planetary disks: linking kepler observables to simulations of late-stage planet formation. *Astrophys. J.* **832** (2016), doi:10.3847/0004-637X/832/1/34.
95. C. D. Dressing *et al.*, Characterizing K2 Candidate Planetary Systems Orbiting Low-Mass Stars III: A High Mass & Low Envelope Fraction for the Warm Neptune K2-55b (2018) (available at http://arxiv.org/abs/1804.05148).
96. H. Mizuno, Formation of the Giant Planets. *Prog. Theor. Phys.* **64**, 544–557 (1980).
97. D. J. Stevenson, Formation of the giant planets. *Planet. Space Sci.* **30**, 755–764 (1982).
98. A.-M. A. Piso, A. N. Youdin, on the minimum core mass for giant planet formation at wide separations. *Astrophys. J.* **786**, 21 (2014).
99. S. L. Li, C. B. Agnor, D. N. C. Lin, Embryo impacts and gas giant mergers. I. dichotomy of jupiter and saturn's core mass. *Astrophys. J.* **720**, 1161–1173 (2010).
100. S.-F. Liu, C. B. Agnor, D. N. C. Lin, S.-L. Li, Embryo impacts and gas giant mergers – II. Diversity of hot Jupiters' internal structure. *Mon. Not. R. Astron. Soc.* **446**, 1685–1702 (2015).
101. A. Bhattacharya *et al.*, WFIRST Exoplanet Mass Measurement Method Finds a Planetary Mass of $39\pm 8 M_\oplus$ for OGLE-2012-BLG-0950Lb (2018) (available at https://arxiv.org/abs/1809.02654).
102. D. Suzuki *et al.*, The exoplanet mass-ratio function from the MOA-II survey: discovery of a break and likely peak at a Neptune mass. *Astrophys. J.* **833**, 145 (2016).
103. E. J. Lee, E. Chiang, Breeding super-earths and birthing super-puffs in transitional disks. *Astrophys. J.* **817**, 90 (2016).
104. M. Ansdell *et al.*, ALMA Survey of Lupus Protoplanetary Disks. I. Dust and Gas Masses. *Astrophys. Journal, Vol. 828, Issue 1, Artic. id. 46, 15 pp. (2016).* **828** (2016), doi:10.3847/0004-637x/828/1/46.
105. M. Ansdell *et al.*, ALMA Survey of Lupus Protoplanetary Disks. II. Gas Disk Radii. *Astrophys. J.* **859**, 21 (2018).
106. S. Seager, M. Kuchner, C. A. Hier-Majumder, B. Militzer, Mass-Radius Relationships for Solid Exoplanets. *Astrophys. J.* **669**, 1279–1297 (2007).
107. L. Zeng, S. Seager, A Computational Tool to Interpret the Bulk Composition of Solid Exoplanets based on Mass and Radius Measurements. *Publ. Astron. Soc. Pacific.* **120**, 983–991 (2008).
108. L. Zeng, D. D. Sasselov, S. B. Jacobsen, Mass-Radius Relation for Rocky Planets Based on PREM. *Astrophys. J.* **819**, 127 (2016).
109. L. Zeng, D. Sasselov, A Detailed Model Grid for Solid Planets from 0.1 through 100 Earth Masses. *Publ. Astron. Soc. Pacific.* **125**, 227–239 (2013).
110. D. C. Swift *et al.*, Mass-Radius Relationships for Exoplanets. *Astrophys. J.* **744**, 59 (2012).
111. R. F. Smith *et al.*, Equation of state of iron under core conditions of large rocky exoplanets. *Nat. Astron.* **2**, 452–458 (2018).
112. F. ~W. Wagner, F. Sohl, H. Hussmann, M. Grott, H. Rauer, Interior structure models of solid exoplanets using material laws in the infinite pressure limit. *Icarus.* **214**, 366–376 (2011).
113. F. ~W. Wagner, N. Tosi, F. Sohl, H. Rauer, T. Spohn, Rocky super-Earth interiors. Structure and internal dynamics of CoRoT-7b and Kepler-10b. *\aap.* **541**, A103 (2012).

114. D. Valencia, D. D. Sasselov, R. J. O'Connell, Detailed Models of Super-Earths: How Well Can We Infer Bulk Properties? *Astrophys. J.* **665**, 1413–1420 (2007).
115. Valencia, D., Ikoma, M., Guillot, T., Nettelmann, N., Composition and fate of short-period super-Earths. *A&A*. **516**, A20 (2010).
116. C. Sotin, O. Grasset, A. Mocquet, Mass-radius curve for extrasolar Earth-like planets and ocean planets. *Icarus*. **191**, 337–351 (2007).
117. O. Grasset, J. Schneider, C. Sotin, A Study of the Accuracy of Mass-Radius Relationships for Silicate-Rich and Ice-Rich Planets up to 100 Earth Masses. *Astrophys. J.* **693**, 722–733 (2009).
118. H. S. Zapolsky, E. E. Salpeter, The Mass-Radius Relation for Cold Spheres of Low Mass. *Astrophys. J.* **158**, 809 (1969).
119. J. J. Fortney, M. S. Marley, J. W. Barnes, Planetary Radii Across Five Orders of Magnitude in Mass and Stellar Insolation: Application to Transits. *Astrophys. J.* **659**, 1661–1672 (2007).
120. C. ~T. Unterborn, E. ~E. Dismukes, W. ~R. Panero, Scaling the Earth: A Sensitivity Analysis of Terrestrial Exoplanetary Interior Models. *ArXiv e-prints* (2015).
121. D. Saumon, G. Chabrier, H. M. van Horn, An Equation of State for Low-Mass Stars and Giant Planets. *Astrophys. J. Suppl. Ser.* **99**, 713 (1995).
122. A. Becker *et al.*, ab initio equations of state for hydrogen (H-REOS.3) and helium (He-REOS.3) and their implications for the interior of brown dwarfs. *Astrophys. J. Suppl. Ser.* **215**, 21 (2014).
123. B. Militzer, W. B. Hubbard, ab initio equation of state for hydrogen-helium mixtures with recalibration of the giant-planet mass-radius relation. *Astrophys. J.* **774**, 148 (2013).
124. M. Chaplin, Water Phase Diagram (2018).
125. M. R. Frank, Y. Fei, J. Hu, Constraining the equation of state of fluid H2O to 80 GPa using the melting curve, bulk modulus, and thermal expansivity of Ice VII. *Geochim. Cosmochim. Acta*. **68**, 2781–2790 (2004).
126. M. French, R. Redmer, Construction of a thermodynamic potential for the water ices VII and X. *Phys. Rev. B*. **91**, 14308 (2015).
127. M. French, T. R. Mattsson, N. Nettelmann, R. Redmer, Equation of state and phase diagram of water at ultrahigh pressures as in planetary interiors. *Phys. Rev. B*. **79**, 54107 (2009).
128. M. Millot *et al.*, Experimental evidence for superionic water ice using shock compression. *Nat. Phys.*, 1 (2018).
129. C. Cavazzoni *et al.*, Superionic and Metallic States of Water and Ammonia at Giant Planet Conditions. *Science (80-. ).* **283**, 44 (1999).
130. J.-A. Hernandez, R. Caracas, Superionic-Superionic Phase Transitions in Body-Centered Cubic H 2 O Ice. *Phys. Rev. Lett.* **117**, 135503 (2016).
131. L. Zeng, D. D. Sasselov, The Effect of Temperature Evolution on the Interior Structure of H2O-rich Planets. *Astrophys. J.* **784**, 96 (2014).
132. "The International Association for the Properties of Water and Steam Revised Release on the IAPWS Formulationd 1995 for the Thermodynamic Properties of Ordinary Water Substance for General and Scientific Use" (2016), (available at http://www.iapws.org.).
133. R. B. Dooley, K. Daucik, "The International Association for the Properties of Water and Steam Revised Release on the Pressure along the Melting and Sublimation Curves of Ordinary Water Substance" (2011), (available at http://www.iapws.org.).

134. W. Wagner, T. Riethmann, R. Feistel, A. H. Harvey, New Equations for the Sublimation Pressure and Melting Pressure of $H_2O$ Ice Ih. *J. Phys. Chem. Ref. Data*. **40**, 43103 (2011).
135. M. D. Knudson *et al.*, Probing the Interiors of the Ice Giants: Shock Compression of Water to 700 GPa and 3.8 g / cm^3. *Phys. Rev. Lett.* **108**, 91102 (2012).
136. E. Sugimura *et al.*, Experimental evidence of superionic conduction in H2O ice. *J. Chem. Phys.* **137** (2012), doi:10.1063/1.4766816.
137. R. Redmer, T. R. Mattsson, N. Nettelmann, M. French, The phase diagram of water and the magnetic fields of Uranus and Neptune. *Icarus*. **211**, 798–803 (2011).
138. E. R. Adams, S. Seager, L. Elkins-Tanton, Ocean Planet or Thick Atmosphere: On the Mass-Radius Relationship for Solid Exoplanets with Massive Atmospheres. *Astrophys. J.* **673**, 1160–1164 (2008).
139. E. D. Lopez, J. J. Fortney, Understanding the Mass-Radius Relation for Sub-neptunes: Radius as a Proxy for Composition. *ApJ.* **792**, 1 (2014).
140. D. Valencia, R. J. O'Connell, D. Sasselov, Internal structure of massive terrestrial planets. *Icarus*. **181**, 545–554 (2006).
141. D. D. Sasselov, M. Lecar, On the Snow Line in Dusty Protoplanetary Disks. *Astrophys. J.* **528**, 995 (2000).
142. T. Sato, S. Okuzumi, S. Ida, On the water delivery to terrestrial embryos by ice pebble accretion. *A&A*. **589** (2016), doi:10.1051/0004-6361/201527069.
143. R. G. Martin, M. Livio, On the evolution of the snow line in protoplanetary discs. *Mon. Not. R. Astron. Soc. Lett.* **425**, L6–L9 (2012).
144. W. Kley, R. P. Nelson, Planet-Disk Interaction and Orbital Evolution. *Annu. Rev. Astron. Astrophys.* **50**, 211–249 (2012).
145. P. J. Armitage, *Astrophysics of Planet Formation* (Cambridge University Press, 2010; http://adsabs.harvard.edu/abs/2010apf..book.....A).
146. S. N. Raymond, C. Cossou, No universal minimum-mass extrasolar nebula: evidence against in situ accretion of systems of hot super-Earths. *Mon. Not. R. Astron. Soc. Lett.* **440**, L11–L15 (2014).
147. S. N. Raymond *et al.*, Planet-planet scattering leads to tightly packed planetary systems. *Astrophys. J.* **696**, L98–L101 (2009).